\documentclass[journal,onecolumn]{IEEEtran}
\usepackage{amsmath,amssymb}
\usepackage{graphicx}
\usepackage{mathrsfs}
\usepackage{tikz}
\usepackage{booktabs}
\usepackage{floatflt}
\usepackage{enumerate}
\usepackage{psfrag}	
\usepackage{array}
\usepackage{multirow,hhline}
\usepackage{exscale}
\usepackage{color}
\usepackage{colortbl}
\usepackage{epsfig}
\usepackage{amsthm}

\newtheorem{theorem}{Theorem}
\newtheorem{lemma}{Lemma}

\begin{document}
\IEEEoverridecommandlockouts

\title{Achieving Net Feedback Gain in the Butterfly Network with a Full-Duplex Bidirectional Relay}
\author{
\IEEEauthorblockN{Anas Chaaban, Aydin Sezgin, and Daniela Tuninetti}
\thanks{%
Anas Chaaban and Aydin Sezgin are with the Ruhr-University of Bochum, 44801 Bochum, Germany, e-mail: anas.chaaban@rub.de, aydin.sezgin@rub.de.

Daniela Tuninetti is with the University of Illinois at Chicago, Chicago, IL 60607 USA, e-mail: danielat@uic.edu.
}
}
\maketitle

\begin{abstract}
A symmetric butterfly network (BFN) with a full-duplex relay operating in a bi-directional fashion for feedback is considered. 
This network is relevant for a variety of wireless networks, including cellular systems dealing with cell-edge users.
Upper bounds on the capacity region of the general memoryless BFN with feedback are derived based on cut-set and cooperation arguments and then specialized to the linear deterministic BFN with really-source feedback. It is shown that the upper bounds are achievable using combinations of the compute-forward strategy and the classical decode-and-forward strategy, thus fully characterizing the capacity region. It is shown that net rate gains are possible in certain parameter regimes.
\end{abstract}

\begin{keywords}
Butterfly Network, Interference Relay Channel with Feedback,
Capacity, Inner bound, Outer Bound.
\end{keywords}

\section{Introduction}
\begin{figure}
\centering
\includegraphics[width=.5\columnwidth]{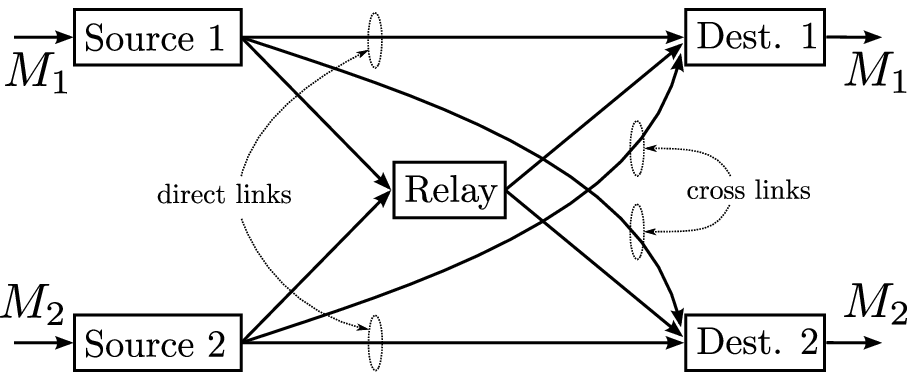}
\caption{The Interference Relay Channel (IRC).}
\label{IRC}
\vspace*{1cm}
\includegraphics[width=.5\columnwidth]{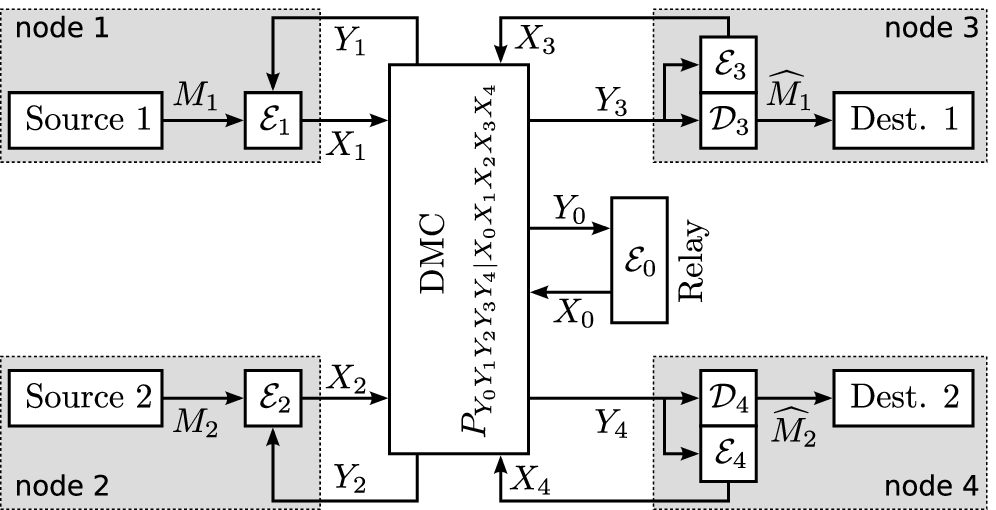}
\caption{The general memoryless Interference Relay Channel with Feedback (IRCF).}
\label{DM_Model}
\vspace*{1cm}
\includegraphics[width=.5\columnwidth]{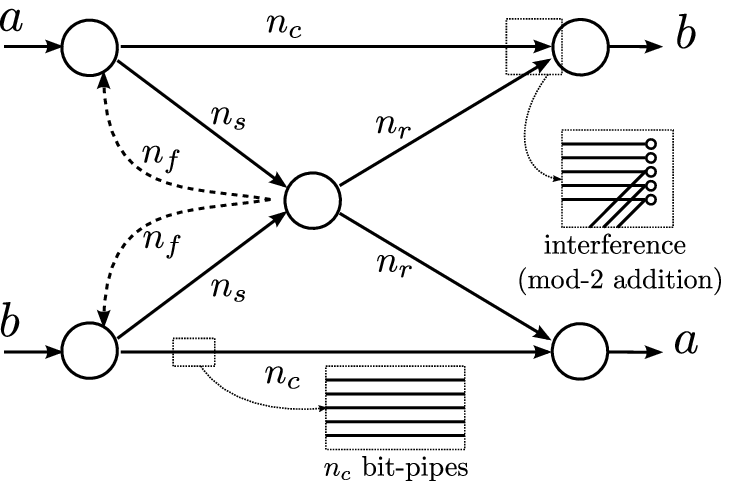}
\caption{The linear deterministic butterfly network with relay-source feedback.}
\label{ldBFNF}
\end{figure}

Ahlswede \cite{Ahlswede} introduced the Interference Channel (IC) as an information theoretic model to capture scenarios 
where simultaneous transmission of dedicated messages by multiple sources to their respective destination takes place on a shared channel. Such a channel is important, for instance, in cellular networks with cell edge users that suffer from interference caused by base stations in neighboring cells.
The phenomenon of interference is not limited to cellular networks and occurs in many other networks such as ad-hoc wireless networks. In the most extreme case, there might be no direct communication link between the transmitting node and its intended receiver due to large obstructing objects. In these cases simply increasing the power level at the transmitting base stations will not resolve the problem. A possible solution is to use dedicated relay stations
to enable communication among source-destination pairs. Such a network was studied by Avestimehr {\em et al.} in \cite{AvestimehrHo} under the assumption that the relay nodes are half-duplex; their channel model is known as {\em the butterfly network (BFN) with a half-duplex relay.}\footnote{Note that the classical butterfly network with multicast message was used by Ahlswede {\em et al.} in \cite{AhlswedeCaiLiYeung} to demonstrate the capabilities of network coding.}
In \cite{AvestimehrHo} the authors exploited network coding ideas in order to design transmission strategies that were shown to be optimal for the linear deterministic approximation of the Gaussian noise BFN at high SNR\footnote{The deterministic approximation of a Gaussian noise network is a deterministic model where the Gaussian additive noises are neglected so as to focus on the interaction of users' signals \cite{AvestimehrDiggaviTse}.} and to achieve capacity to within 1.95~bits per channel use at any finite SNR. Note that the BFN is a special case of the interference relay channel (IRC) \cite{SahinErkip,MaricDaboraGoldsmith_IT,ChaabanSezgin_IT_IRC} shown in Fig. \ref{IRC} obtained by setting the direct links to zero.
In this paper, we consider a BFN in which the nodes are full-duplex and where a dedicated feedback channel exists from the relay to the sources. From a slightly different perspective, the resulting setup can be considered as an IC utilizing a bi-directional relay for interference management to achieve higher data rates.

\subsection{Contributions}
The main contribution of this paper is the characterization of the capacity region of the full-duplex linear deterministic BFN with relay-source feedback.

First, we introduce the general memoryless IRC with Feedback (IRCF) where each node is full-duplex and has both an input to and an output from the channel. For such an IRCF, depicted in Fig.~\ref{DM_Model}, we provide upper bounds on the achievable rates based on the cut-set bound \cite[Thm.15.10.1]{CoverThomas} and based on an upper bound recently derived for the general cooperative IC \cite{Tuninetti_ITW}. We then specialize these upper bounds to the linear deterministic BFN with relay-source feedback depicted in Fig.~\ref{ldBFNF}
for which we provide a complete characterization of the capacity region.

Our achievable strategies aim to establish cooperation among the source nodes and the relay and to exploit the feedback from the relay to the source nodes. The relay participates in the delivery of the messages, since clearly in the setup of Fig.~\ref{ldBFNF} communications is only possible via the relay. 
We develop transmission strategies where both the relay-destination links and the feedback links are used to deliver messages from the sources to the destination. We use the following main ingredients:
\begin{itemize}

\item \textbf{Decode-forward (DF)}: 
Each source sends a ``D-signal'' to be decoded and forwarded by the relay using classical DF~\cite{CoverElgamal}.

\item \textbf{Compute-forward (CF)}:
Each source sends a ``C-signal''. The relay decodes a function (in our specific case the sum) of the C-signals and forwards it to the destinations. Since the processing at the relay does not involve decoding each C-signal separately, but ``computing'' their sum, the strategy is refered to as compute-forward \cite{NazerGastpar}. This strategy is designed in such a way that each destination can decode both the interfering C-signal and the forwarded sum of C-signals. Backward decoding is used at the destinations to recover the desired C-signal. A similar strategy was used in \cite{ChaabanSezgin_IT_IRC} for the IRC, and a half-duplex variant was also used in \cite{AvestimehrHo} for the half-duplex BFN.

\item \textbf{Cooperative Neutralization (CN)}:
Each source sends two ``N-signals'': a ``present N-signal'' and a ``future N-signal''. The future N-signal is intended for the relay only, which computes the sum of the future N-signals. The relay then forwards this sum in the next channel use (note that the ``future N-signals'' of the $i$-th channel use are ``present N-signals'' in channel use $i+1$). This strategy is designed as follows. The forwarded N-signal sum from the relay and the interfering N-signal from the cross link interfere at the destination in such a way that neutralizes interference (on the fly) leaving the desired N-signal interference free. A similar strategy was used for the interference channel with cooperation in \cite{YangTuninentti_Asilomar} and for the half-duplex BFN in \cite{AvestimehrHo}.

\item \textbf{Feedback (F)}:
Each source sends an ``F-signal'' to the relay. The two sources and the relay operate on the F-signals as in the bi-directional relay channel \cite{RankovWittneben,KimDevroyeMitranTarokh,AvestimehrSezginTse}. In a nutshell, the bi-directional relay channel is a setup consisting of two nodes that want to establish two way communications via a relay node, where each node is a transmitter and a receiver at the same time. In the BFN with feedback, the relay-source feedback channels together with the source-relay forward channels establish such bi-directional relay channel. Therefore, as in the bi-directional relay channel, each source is able to obtain the F-signal of the other source. Then, the sources use their cross link to deliver the F-signal of the other source node to its respective destination.

\end{itemize}
Our general achievable strategy uses a combination of these techniques depending on the channel parameters.
The following give a rational as of why certain schemes should be used for a specific scenarios:
\begin{itemize}

\item
If the source-relay channel is stronger than the source-destination (cross) channel, then the sources can pass some future information to the relay without the destinations noticing (below their noise floor). This future information is to be used in the next channel use for interference neutralization. If the source-relay channel is weaker than the source-destination channel, the CN strategy should be avoided since the transmission of future information to the relay disturbs the destinations in this case.

\item
On the other hand, the F strategy is to be used when the source-destination (cross) channel is stronger than the relay-destination channel. In this case, the sources can send the signal acquired via feedback to the destinations, which is received by the destination at a higher SNR than the relay signal. This allows the destination to decode this signal, strip it, and then proceed with decoding the relay signal. Otherwise, if the cross channel is weaker than the relay-destination channel, then such transmission would disturb the relay transmission and should be avoided.

\item
In the CF strategy, each destination has to decode two observations of the C-signals in each channel use (the interfering C-signal and the sum of the C-signals), whereas the relay has to decode only one observation (the C-signal sum). Therefore, this scheme requires more levels at the destinations than at the relay. For this reason, the CF strategy is to be used by the relay if the source-relay channel is weaker than either the relay-destination channel or the source-destination (cross) channel (as in \cite{ChaabanSezgin_IT_IRC}).

\item
The DF strategy can be always used to achieve asymmetric rate points.

\end{itemize}
By using this intuition, we design achievable strategies for different parameter regimes that meet the derived outer bounds for the linear deterministic BFN with relay-source feedback, thus characterizing its capacity region completely.

\subsection{Paper Organization}
The general flow of the paper is as follows. 
We define the general memoryless IRCF in Sect. \ref{DMIRCWF} where we also provide upper bounds.
The linear deterministic BFN with relay-source feedback is defined in Sect. \ref{Section:LDBFNF} and its upper bounds are derived in Sect. \ref{Section:UpperBoundsfortheld-BFNF}. The coding strategies (DF, CF, CN, and F) that constitute the basic building block of our achievable schemes are described in Sect. \ref{Section:CodingStrategies}. The capacity achieving scheme is described and analyzed in Sect. \ref{LowerBoundNcNr} and \ref{LowerBoundNrNc}, for the two regimes where relay-source feedback does not and does, respectively, increase the capacity with respect to the non-feedback case. We discuss the net-gain due to feedback in Sect. \ref{Section:NetGain}. Sect. \ref{summary} concludes the paper.

\subsection{Notation}
We use $X^N$ to denote the length-$N$ sequence $(X_1,X_2,\dots,X_N)$, $(x)^+ := \max\{0,x\}$ for $x\in\mathbb{R}$, and $0_\ell$ to denote the all-zero vector of length $\ell\in\mathbb{N}$. For a vector $x(i)$ given as $$x(i)=\left[\begin{array}{c}x^{[1]}(i)\\x^{[2]}(i)\\\vdots\\x^{[K]}(i)\end{array}\right],$$ $i$ denotes the time index, and $x^{[k]}(i)$ is the $k$-th component of $x(i)$, which can be scalar or vector depending on the context.
$x^T$ is the transpose of the vector $x$.

\section{The memoryless IRC with relay-source feedback: channel model and outer bounds}
\label{DMIRCWF}
In Section~\ref{DMIRCWF:ch model} we introduce the memoryless IRC with general feedback even though in the rest of the paper we will be analyzing the case of relay-source feedback only. The reason for doing so is that the general feedback model allows us to easily describe the proposed outer bounds for the relay-source feedback model in Section~\ref{DMIRCWF:upper}.

\subsection{The memoryless IRC with general feedback}
\label{DMIRCWF:ch model}

A memoryless IRC with general feedback is a five node network with a relay (node 0), two sources (nodes 1 and 2), and two destinations (nodes 3 and 4) sharing the same channel, as shown in Fig.~\ref{DM_Model}. All nodes are full-duplex and causal. 
Node~$j$, $j\in\{1,2\}$, has an independent message $M_j\in\{1,\dots,2^{N R_j}\}$, where $N\in\mathbb{N}$ is the code-length and $R_j\in\mathbb{R}_+$ the rate in bits per channel use, to be sent to  node~$j+2$. 
The operations performed at each node can be described in general as follows:
\begin{itemize}
\item Node 0 receives $Y_0$ and sends $X_0$, where the $i$-th symbol of $X_0^N$ is constructed from $Y_0^{i-1}$ using an encoding function $\mathcal{E}_{0,i}$, i.e., $X_{0,i}=\mathcal{E}_{0,i}(Y_0^{i-1})$.
\item Node 1 receives feedback information $Y_1$ and sends $X_1$, where $X_{1,i}$ is constructed from the message $M_1$ and from $Y_1^{i-1}$ using an encoding function $\mathcal{E}_{1,i}$, i.e., $X_{1,i}=\mathcal{E}_{1,i}(M_1,Y_1^{i-1})$.
\item Node 2 operates similarly to node 1, i.e., $X_{2,i}=\mathcal{E}_{2,i}(M_2,Y_2^{i-1})$.
\item Node 3 receives $Y_3$ and sends $X_3$, where $X_{3,i}$ is constructed from $Y_3^{i-1}$ using an encoding function $\mathcal{E}_{3,i}$, i.e., $X_{3,i}=\mathcal{E}_{3,i}(Y_3^{i-1})$. After $N$ channel uses, node~3/destination~1 tries to obtain $M_1$ from $Y_3^N$ using a decoding function $\mathcal{D}_{3}$, i.e., $\widehat{M}_1=\mathcal{D}_3(Y_3^N)$. An error occurs if $M_1\neq\widehat{M}_1$.
\item Node 4 operates similarly to node~3/destination~1, i.e., $X_{4,i}=\mathcal{E}_{4,i}(Y_4^{i-1})$ and $\widehat{M}_2=\mathcal{D}_4(Y_4^N)$. An error occurs if $M_2\neq\widehat{M}_2$.
\end{itemize}
The channel has transition probability $P_{Y_0,Y_1,Y_2,Y_3,Y_4|X_0,X_1,X_2,X_3,X_4}$ and is assumed to be memoryless, that is, for all $i\in\mathbb{N}$ the following Markov chain holds
\begin{align*}
&(W_1,W_2,
X_0^{i-1},X_1^{i-1},X_2^{i-1},X_3^{i-1},X_4^{i-1},
Y_0^{i-1},Y_1^{i-1},Y_2^{i-1},Y_3^{i-1},Y_4^{i-1})
\\&\to (X_{0,i},X_{1,i},X_{2,i},X_{3,i},X_{4,i})
   \to (Y_{0,i},Y_{1,i},Y_{2,i},Y_{3,i},Y_{4,i}).
\end{align*}
We use the standard information theoretic definition of a code, probability of error and achievable rates \cite{CoverThomas}. We aim to characterize the capacity defined as the convex closure of the set of non-negative rate pairs $(R_1,R_2)$ such that $\max_{j\in\{1,2\}}\mathbb{P}[M_j\neq\widehat{M}_j] \to 0$ as $N\to\infty$.

This model generalizes various well studied channel models. For instance, it models the classical IC \cite{Carleial} (for $Y_1=Y_2=Y_0=X_0=X_3=X_4=\emptyset$), the IC with cooperation \cite{Tuninetti_ITW} (for $Y_0=X_0=\emptyset$), the classical IRC \cite{MaricDaboraGoldsmith_IT,ChaabanSezgin_IT_IRC} (for $Y_1=Y_2=X_3=X_4=\emptyset$), etc.

\subsection{Upper bounds for the memoryless IRC with relay-source feedback}
\label{DMIRCWF:upper}
The memoryless IRC with relay-source feedback is obtained from the model in Section~\ref{DMIRCWF:ch model} by setting $X_3=X_4=\emptyset$. We next derive several upper bounds on achievable rate pairs for the general memoryless IRC with relay-source feedback. We note that the described techniques apply to the general IRCF and do not require necessarily $X_3=X_4=\emptyset$.
We start with the cut-set bound \cite{CoverThomas} 
and then we adapt upper bounds for the general memoryless IC with cooperation given in \cite{Tuninetti_ITW} to our channel model.

\subsubsection{Cut-set bounds}
\label{DMIRCWF:upper:cutset}

The cut-set bound \cite{CoverThomas} applied to a general network with independent messages at each node states that an achievable rate vector must satisfy
\begin{align}
R(\mathcal{S}\to\mathcal{S}^c) \leq I(X(\mathcal{S}); Y(\mathcal{S}^c)|X(\mathcal{S}^c)),
\label{eq:general cuset}
\end{align}
for some joint distribution on the inputs, where $\mathcal{S}$ is a subset of the nodes in the network,  $\mathcal{S}^c$ is the complement of $\mathcal{S}$, and $R(\mathcal{S}\to\mathcal{S}^c)$ indicates the sum of the rates from the source nodes in $\mathcal{S}$ to the destination nodes in $\mathcal{S}^c$.

For the IRCF, by using~\eqref{eq:general cuset}, the rate $R_1$ can be bounded as
\begin{subequations}
\begin{align}
\label{CutSetBound1}
R_1&\leq I(X_1;Y_0,Y_2,Y_3|X_0,X_2)\\
\label{CutSetBound2}
R_1&\leq I(X_1,X_2;Y_0,Y_3|X_0)\\
\label{CutSetBound3}
R_1&\leq I(X_0,X_1;Y_2,Y_3|X_2)\\
\label{CutSetBound4}
R_1&\leq I(X_0,X_1,X_2;Y_3),
\end{align}
\label{CutSetBoundAllR1}
\end{subequations}
for some input distribution $P_{X_0,X_1,X_2}$. 

Similarly, we can bound $R_2$ by replacing the subscripts 1, 2, and 3 with 2, 1, and 4, respectively, in~\eqref{CutSetBoundAllR1}.

The sum-rate can be bounded as
\begin{subequations}
\begin{align}
\label{CutSetBound5}
R_1+R_2&\leq I(X_1,X_2;Y_0,Y_3,Y_4|X_0)\\
\label{CutSetBound6}
R_1+R_2&\leq I(X_0,X_1,X_2;Y_3,Y_4),
\end{align}
\label{CutSetBoundAllRsum}
\end{subequations}
for some input probability distribution $P_{X_0,X_1,X_2}$.

\subsubsection{Cooperation upper bounds}
\label{DMIRCWF:upper:dt}

As mentioned earlier, the IC with general cooperation is a special case of the IRCF obtained by setting $Y_0=X_0=\emptyset$. An upper bound for the sum-capacity of the IC with general cooperation is \cite{Tuninetti_ITW}
\begin{subequations}
\begin{align}
R_1+R_2  &\leq  I(X_1;Y_3,Y_2| Y_4,X_2,X_3,X_4) + I(X_1,X_2,X_3;Y_4|X_4),\\
R_1+R_2  &\leq  I(X_2;Y_4,Y_1| Y_3,X_1,X_3,X_4) + I(X_1,X_2,X_4;Y_3|X_3). 
\end{align}
\label{eq:tuninetti itw 2012 lausanne}
\end{subequations}
for some $P_{X_1,X_2,X_3,X_4}$. 

\begin{subequations}
In the interference relay channel with feedback, if we let the relay perfectly cooperate with one of the other nodes in the network, then the model again reduces to an IC with general cooperation in which one of the nodes has an enhanced input and output. Since cooperation cannot decrease capacity, any outer bound for the IC with general cooperation is an upper bound to the capacity of the interference relay channel with feedback. In particular, if node~$j$, $j\in\{1,2,3,4\}$, cooperates with the relay (node~0), then in~\eqref{eq:tuninetti itw 2012 lausanne} we replace $X_j$ with $(X_j,X_0)$ and $Y_j$ with $(Y_j,Y_0)$. Moreover, since we do not consider feedback from the destinations in this paper, we set $X_3=X_4=\emptyset$ after this substitution.
This yields the following upper bounds:
\begin{enumerate}
\item Full cooperation between node~1 and node~0, giving an IC with bi-directional cooperation between nodes 1 and 2 where node 1 sends $(X_1,X_0)$ and receives $(Y_1,Y_0)$: 
\begin{align}
R_1+R_2  &\leq  I(X_1,X_0;Y_3,Y_2| Y_4,X_2) + I(X_1,X_0,X_2;Y_4),    \label{eq:tuninetti itw 2012 lausanne 1-1}\\
R_1+R_2  &\leq  I(X_2;Y_4,Y_1,Y_0| Y_3,X_1,X_0) + I(X_1,X_0,X_2;Y_3).\label{eq:tuninetti itw 2012 lausanne 1-2}
\end{align}
\item Full cooperation between node~2 and node~0, giving an IC with bi-directional cooperation between nodes 1 and 2 where node 2 sends $(X_2,X_0)$ and receives $(Y_2,Y_0)$:
\begin{align}
R_1+R_2  &\leq  I(X_1;Y_3,Y_2,Y_0| Y_4,X_2,X_0) + I(X_1,X_2,X_0;Y_4),\label{eq:tuninetti itw 2012 lausanne 2-1}\\
R_1+R_2  &\leq  I(X_2,X_0;Y_4,Y_1| Y_3,X_1) + I(X_1,X_2,X_0;Y_3).    \label{eq:tuninetti itw 2012 lausanne 2-2}
\end{align}
\item Full cooperation between node~3 and node~0, giving an IC with uni-directional cooperation between node 3 and 4 and with feedback from node 3 to nodes 1 and 2, where node 3 sends $X_0$ and receives $(Y_3,Y_0)$: 
\begin{align}
R_1+R_2  &\leq  I(X_1;Y_3,Y_0,Y_2| Y_4,X_2,X_0) + I(X_1,X_2,X_0;Y_4),    \label{eq:tuninetti itw 2012 lausanne 3-1}\\
R_1+R_2  &\leq  I(X_2;Y_4,Y_1| Y_3,Y_0,X_1,X_0) + I(X_1,X_2;Y_3,Y_0|X_0).\label{eq:tuninetti itw 2012 lausanne 3-2}
\end{align}
\item Finally, full cooperation between node~4 and node~0, giving an IC with uni-directional cooperation between node 4 and 3 and with feedback from node 4 to nodes 1 and 2, where node 4 sends $X_0$ and receives $(Y_4,Y_0)$:
\begin{align}
R_1+R_2  &\leq  I(X_1;Y_3,Y_2| Y_4,Y_0,X_2,X_0) + I(X_1,X_2;Y_4,Y_0|X_0),\label{eq:tuninetti itw 2012 lausanne 4-1}\\
R_1+R_2  &\leq  I(X_2;Y_4,Y_0,Y_1| Y_3,X_1,X_0) + I(X_1,X_2,X_0;Y_3).    \label{eq:tuninetti itw 2012 lausanne 4-2}
\end{align}
\end{enumerate}

\label{eq:tuninetti itw 2012 lausanne allall}
\end{subequations}

These upper bounds will be used next to upper bound the capacity region of the butterfly network with relay-source feedback. As it turns out, these bounds suffice to characterize the capacity of the symmetric linear deterministic butterfly network.

\section{The linear deterministic butterfly network with feedback} 
\label{Section:LDBFNF}

We consider here a special case for the IRC with relay-source feedback described in the previous section, namely the linear deterministic channel that is by now customarily used to approximate a Gaussian noise network at high SNR as originally proposed by \cite{AvestimehrDiggaviTse}.
 
We assume a dedicated out-of-band feedback channel between node 0 on one side, and nodes 1 and 2 on the other side. For this reason, we write $X_0$ as $(X_r,X_f)$ where $X_r$ is the in-band relay signal to the destinations and $X_f$ is the out-of-band feedback signal to the sources. 
The input-output relations of this linear deterministic IRCF with out-of-band feedback from the relay to the sources is\begin{subequations}
\begin{align}
Y_0&= \mathbf{S}^{q-n_{10}} X_1 + \mathbf{S}^{q-n_{20}} X_2,\\
Y_1&= \mathbf{S}^{q-n_{01}} X_f,\\
Y_2&= \mathbf{S}^{q-n_{02}} X_f,\\
Y_3&= \mathbf{S}^{q-n_{13}} X_1 + \mathbf{S}^{q-n_{23}} X_2 + \mathbf{S}^{q-n_{03}} X_r,\\
Y_4&= \mathbf{S}^{q-n_{14}} X_1 + \mathbf{S}^{q-n_{24}} X_2 + \mathbf{S}^{q-n_{04}} X_r,
\end{align}
\label{eq:generalchannelmodel}
\end{subequations}
where $Y_0$ is the channel output at relay, $Y_1$ and $Y_2$ are the received feedback signal at the sources, and $Y_3$ and $Y_4$ are the received signals at the destinations. Here $q := \max\{n_{jk}\}$, with $n_{jk}\in\mathbb{N}$ for $j\in\{0,1,2\}$ and $k\in\{0,1,2, 3,4\}$ and $\mathbf{S}$ is the $q\times q$ shift matrix
\begin{align*}
\mathbf{S}
:=
\begin{bmatrix}
0 & 0 & 0 &\dots\\
1 & 0 & 0 &\dots\\
0 & 1 & 0 &\dots\\
\vdots&\vdots&\vdots&\ddots\\
\end{bmatrix}.
\end{align*}
All signals are binary vectors of length $q$ and addition is the component-wise addition over the binary field.

As the number of parameters in the general channel model in \eqref{eq:generalchannelmodel} is large, we resort to a symmetric setup for simplicity of exposition. This simplification reduces the number of parameters, and thus leads to complete analytical, clean, and insightful capacity region characterization. In the {\em symmetric} scenario 
the channel model in \eqref{eq:generalchannelmodel} has the following parameters
\begin{align*}
  &n_{13}=n_{24}=0   &&\quad\text{(direct channel),}
\\&n_{14}=n_{23}=n_c &&\quad\text{(cross channel),}
\\&n_{03}=n_{04}=n_r &&\quad\text{(relay-destination channel),}
\\&n_{10}=n_{20}=n_s &&\quad\text{(source-relay channel),}
\\&n_{01}=n_{02}=n_f &&\quad\text{(feedback channel).}
\end{align*}
Thus, the symmetric linear deterministic BFN with feedback shown in Fig.~\ref{ldBFNF} has the following input-output relationship
\begin{subequations}
\begin{align}
Y_0&= \mathbf{S}^{q-n_{s}} \big(X_1 + X_2\big),\\
Y_1&= \mathbf{S}^{q-n_{f}} X_f,\\
Y_2&= \mathbf{S}^{q-n_{f}} X_f,\\
Y_3&= \mathbf{S}^{q-n_{c}} X_2 + \mathbf{S}^{q-n_{r}} X_r,\\
Y_4&= \mathbf{S}^{q-n_{c}} X_1 + \mathbf{S}^{q-n_{r}} X_r.
\end{align}
\label{eq:finallythemodelwestudy}
\end{subequations}

The main focus of the rest of the paper is to determine the capacity region of the network described by~\eqref{eq:finallythemodelwestudy}. In the following section, we provide matching upper and lower bounds for the linear deterministic BFN with feedback thereby completely characterizing the capacity region.

\section{Upper Bounds for the linear deterministic BFN with feedback}
\label{Section:UpperBoundsfortheld-BFNF}

In this section we specialize the general bounds given in Section \ref{DMIRCWF:upper} to the linear deterministic BFN described in Section~\ref{Section:LDBFNF}. Our main result is as follows.
\begin{theorem}
\label{Theorem:CapacityOuter}
The capacity region of the linear deterministic BFN with source-relay feedback is contained in the set of rate pairs $(R_1,R_2)$ such that
\begin{subequations}
\begin{align}
\label{UpperBound1}
0\leq R_1&\leq \min\{n_s,n_r+n_f,\max\{n_c,n_r\}\}\\
\label{UpperBound2}
0\leq R_2&\leq \min\{n_s,n_r+n_f,\max\{n_c,n_r\}\}\\
\label{UpperBound3}
R_1+R_2&\leq\max\{n_r,n_c\}+n_c\\
\label{UpperBound4}
R_1+R_2&\leq\max\{n_r,n_c\}+(n_s-n_c)^+\\
\label{UpperBound5}
R_1+R_2&\leq n_s+n_c.
\end{align}
\label{eq:UpperBound:Theorem:CapacityOuter}
\end{subequations}
\end{theorem}
The details of the proof can be found in the Appendix.

An intuitive explanation of the single-rate bounds in Thm.~\ref{Theorem:CapacityOuter} is as follows.

Since communications is only possible via the relay, source 1 can not send more bits per channel use than the relay can receive; thus, we have the bound $R_1\leq n_s$ in~\eqref{UpperBound1}. Now assume that the channel to the relay is very strong (say of infinite capacity); in this case, the rate achieved by a source can not exceed the capacity of the outgoing channels from the relay, i.e., $n_r+n_f$ in~\eqref{UpperBound1}. Finally, the rate $R_1$ can not exceed the amount of information that can be received by node~3/destination~1, which is given by $\max\{n_c,n_r\}$, and hence the bound $R_1\leq\max\{n_c,n_r\}$  in~\eqref{UpperBound1}. Similar reasoning holds for the bound in~\eqref{UpperBound2}.

Interestingly, the sum-rate bounds in~\eqref{UpperBound3}-\eqref{UpperBound5} do not depend on the feedback parameter $n_f$. As we shall see in the following sections, given $n_r>n_c$, the region in~\eqref{eq:UpperBound:Theorem:CapacityOuter} is as for $n_f=0$, i.e., no gain from the availability of a dedicated relay-source feedback channel. In this case, the relay-destination link is so strong that the relay can help the destinations resolve their signals without the need of source cooperation. On the other hand, when $n_r<n_c$, relaying can be improved upon by source cooperation enabled by the presence of feedback; in this case, we can have a `net-gain' from feedback that is larger than the `cost' of feedback. We will expand on this idea after we proved the achievability of the outer bound in Thm.~\ref{Theorem:Capacity}.

\section{Achievable Strategies}
\label{Section:CodingStrategies}

The main result of this section is as follows:
\begin{theorem}
\label{Theorem:Capacity}
The outer bound region in Thm.~\ref{Theorem:CapacityOuter} is achievable.
\end{theorem}

Before we prove the achievability of the outer bound in Thm.~\ref{Theorem:CapacityOuter}, we describe the different coding strategies that we will use in the achievability proof. Each strategy is discussed separately in the rest of this section.
The proof of Thm.~\ref{Theorem:Capacity} is a careful combination of these strategies for different parameter regimes.
The actual proof of Thm.~\ref{Theorem:Capacity}, due to its length, is split between Section~\ref{LowerBoundNcNr} and Section~\ref{LowerBoundNrNc}.

\subsection{Cooperative interference neutralization}
\label{Section:CodingStrategies:CN}
We propose a signaling scheme which we call {\em cooperative interference neutralization}, or CN for short. The main idea of CN is to allow the relay to know some information about future source transmissions, in order to facilitate interference neutralization. This is done as follows. Each source sends two N-signals in the $i$-th channel use, which we call $u_{j,n}(i)$ and $u_{j,n}(i+1)$, where $j\in\{1,2\}$ is the source index, the subscript $n$ is used to denote N-signals, and where $i$ is the channel use index. $u_{j,n}(i)$ is the N-signal to be decoded by the destination in the $i$-th channel use, while $u_{j,n}(i+1)$ is to be decoded in the next channel use $i+1$. Therefore, the source sends the present and the future N-signals. The future one, $u_{j,n}(i+1)$ is intended for the relay, and is not decoded at the destinations. The relay attempts to decode $u_{1,n}(i+1)\oplus u_{2,n}(i+1)$ in the $i$-th channel use. This sum is then sent in the next channel use $i+1$, on the same levels where $u_{2,n}(i+1)$ is observed at node~3/destination~1 (note that $u_{2,n}(i+1)$ is interference from node~3/destination~1's perspective), resulting in interference neutralization since $u_{1,n}(i+1)\oplus u_{2,n}(i+1)\oplus u_{2,n}(i+1)=u_{1,n}(i+1)$. This allows node~3/destination~1 to decode its desired N-signal in channel use $i+1$.

\begin{figure}[ht]
\centering
\includegraphics[width=.45\textwidth]{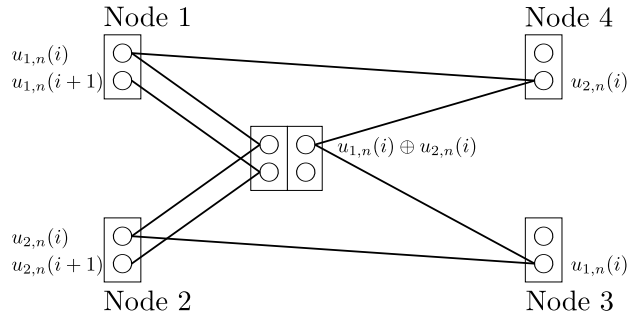}
\caption{A graphical illustration of the CN strategy. Due to interference neutralization, node~3/destination~1 receives $u_{2,n}(i)\oplus u_{1,n}(i)\oplus u_{2,n}(i)=u_{1,n}(i)$ interference free at the second level. Similarly, node~4/destination~2 obtains $u_{2,n}(i)$. Using this strategy in this setup, each source can send 1 bit per channel use. Note how the sources pass the future N-signals to the relay without disturbing the destinations.}
\label{EG_CN}
\end{figure}

In Fig.~\ref{EG_CN}, as well as in similar figures in the following, the vertical bars represent bit vectors and the circles inside them represent bits. On the left we represent the bits of the sources (node~1 on top and node~2 at the bottom) and on the right the bits of the destinations (node~4/destination~2 on top and node~3/destination~1 at the bottom); the relay is represented in the middle. Lines connecting circles represent bit-pipes, and when a level (circle) receives 2 bit-pipes (lines), the modulo-2 sum of the bits is observed (valid at the relay and the destinations). The in-band channel is drawn in black, while the out-of-band feedback channel is drawn in red. When the red channel is not shown, this means that either $n_f=0$ (no feedback channel to the sources) or the feedback channel is not used. 

An illustrative example for CN is given in Fig.~\ref{EG_CN}. In Fig.~\ref{EG_CN} destination~1/node~3 receives on its the second level $u_{2,n}(i)\oplus u_{1,n}(i)\oplus u_{2,n}(i)=u_{1,n}(i)$, that is, thanks to CN, the signal $u_{1,n}(i)$ is received interference free.  Similarly, destination~2/node~4 obtains $u_{2,n}(i)$ interference free.  Note how the sources pass the future N-signals to the relay without disturbing the destinations.

From Fig. \ref{EG_CN} we remark that by using CN each source can send $R_n$ bits per channel use over $R_n$ levels at the destination while using $2R_n$ levels at the relay. Due to this fact, this strategy is preferable when $n_s$ is larger than $n_c$.

To realize the CN strategy we use block Markov coding. Each source sends $N$ signals in $N+1$ channel uses. Starting with an initialization step, the sources send $u_{j,n}(1)$ in channel use $i=0$ while the relay remains silent. Then, each source sends both $u_{j,n}(i)$ and $u_{j,n}(i+1)$ in the $i$-th channel use for $i=1,\dots,N-1$ while the relay sends $u_{1,n}(i)\oplus u_{2,n}(i)$. Finally, in the $N$-th channel use, each source sends $u_{j,n}(N)$ only and the relay sends $u_{1,n}(N)\oplus u_{2,n}(N)$. Each destination decodes its desired N-signal starting from $i=1$ till $i=N$. Thus, assuming that $u_{j,n}$ is a binary vector of length $R_n$, each source is able to successfully deliver $NR_n$ bits over the span of $N+1$ channel uses. Hence, the rate per channel use would be $\frac{N}{N+1}R_n$ which approaches $R_n$ for large $N$. This factor $\frac{N}{N+1}$ will be ignored from now on, as we always choose $N$ to be large.

A strategy similar to the CN strategy was also used in the interference channel with generalized feedback in \cite{YangTuninentti_Asilomar}, where the sources exchanges bits below the noise floor of the receivers, which are then used in the next slot to `zero force' the interference. A half-duplex variant of this scheme was also used for the half-duplex BFN in \cite{AvestimehrHo}.

\subsection{Compute-forward}
\label{Section:CodingStrategies:CF}
\begin{figure}[ht]
\centering
\includegraphics[width=.55\textwidth]{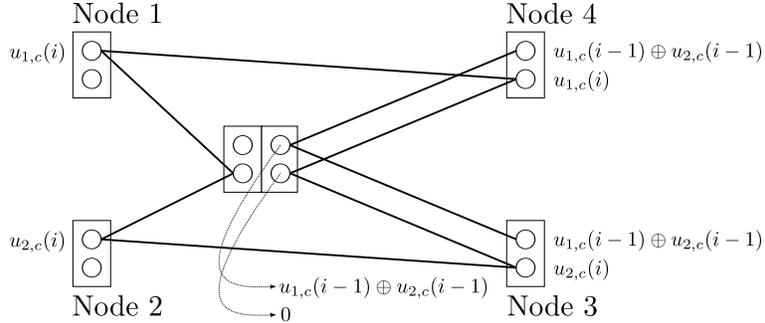}
\caption{A graphical illustration of the CF strategy. Node~3/destination~1 decodes $u_{1,c}(i-1)\oplus u_{2,c}(i-1)$ and $u_{2,c}(i)$ in the $i$-th channel use. By backward decoding, it can add $u_{1,c}(i-1)\oplus u_{2,c}(i-1)$ (decoded in the $i$-th channel use and $u_{2,c}(i-1)$ (decoded in channel use $i-1$) to obtain its desired CP signal $u_{1,c}(i-1)$. Using this strategy in this setup, each source can send 1 bit per channel use.}
\label{EG_CF}
\end{figure}

We use compute-forward at the relay \cite{NazerGastpar} (CF) to deliver both source messages to both destinations. The CF strategy works as follows (see Fig. \ref{EG_CF} for an example). Each source sends a signal $u_{j,c}(i)$ in the $i$-th channel use, $i=1,\dots,N$ and where we use the subscript $c$ to indicate C-signals. The relay decodes the function/sum $u_{1,c}(i)\oplus u_{2,c}(i)$ in the $i$-th channel use and sends it in the next channel use on a different level at the destinations. This process is repeated from $i=1$ till $i=N+1$, where the sources are active in channel uses $i=1,\dots,N$ and the relay is active in channel uses $i=2,\dots,N+1$.

Thus, node~3/destination~1 for instance receives $u_{2,c}(i)$ and $u_{1,c}(i-1)\oplus u_{2,c}(i-1)$ in the $i$-th channel use, $i=2,\dots,N$. In the first channel use, it only receives $u_{2,c}(1)$ since the relay has no information to send in this channel use. In channel use $N+1$, it only receives $u_{1,c}(N)\oplus u_{2,c}(N)$ from the relay since the sources do not send in this channel use. Decoding is performed backwards starting from $i=N+1$, where only the relay is active and thus $u_{1,c}(N)\oplus u_{2,c}(N)$ is decoded. In the $N$-th channel use, destination 1 decodes $u_{1,c}(N-1)\oplus u_{2,c}(N-1)$ and $u_{2,c}(N)$. Then, it adds the two observations of the signals with time index $N$, i.e., $u_{1,c}(N)\oplus u_{2,c}(N)$ and $u_{2,c}(N)$ to obtain its desired signal $u_{1,c}(N)$. Similar decoding is performed at node~4/destination~2. Decoding proceeds backwards till $i=1$ is reached. If the signals $u_{j,c}$ are binary vectors of length $R_c$, then each source achieves $R_c$ bits per channel use for large $N$ using this strategy. The signals sent using this strategy are ``public'', in the sense of the Han and Kobayashi's achievable region for the classical IC \cite{HanKobayashi}, i.e., each destination decode both C-signals from source 1 and 2.

An example of CF strategy is given in Fig.~\ref{EG_CF}. Here node~3/destination~1 decodes $u_{1,c}(i-1)\oplus u_{2,c}(i-1)$ and $u_{2,c}(i)$ in the $i$-th channel use. By backward decoding, it can add $u_{1,c}(i-1)\oplus u_{2,c}(i-1)$ (decoded in the $i$-th channel use) and $u_{2,c}(i-1)$ (decoded in channel use $i-1$) to obtain its desired signal $u_{1,c}(i-1)$.

Notice from Fig. \ref{EG_CF} that the CF strategy allows the sources to send $R_c$ bits each while using $R_c$ levels at the relay and $2R_c$ levels at the destinations. For this reason, this strategy is preferable when the number of levels at the destinations $\max\{n_c,n_r\}$ is larger than $n_s$.

\subsection{Feedback}
\label{Section:CodingStrategies:F}

\subsubsection{Symmetric} 
Here both sources use the same strategy. 
This strategy exploits the feedback channel between the relay and the sources to establish cooperation between the sources. It is similar to the scheme used in the linear deterministic bi-directional relay channel in  \cite{NarayananPravinSprintson}. Each source~$j$, $j\in\{1,2\}$, sends a feedback (F) signal $u_{j,f}(i)$ in the $i$-th channel use, where the subscript $f$ is used to indicate F-signals. The relay decodes the sum $u_{1,f}(i)\oplus u_{2,f}(i)$ in the $i$-th channel use and feeds it back to the sources in channel use $i+1$. In channel use $i+1$, source 1 for instance decodes $u_{1,f}(i)\oplus u_{2,f}(i)$ from the feedback channel, and extracts $u_{2,f}(i)$, having its own signal $u_{1,f}(i)$ as ``side information''. Then, it sends this information to destination 2 using its cross channel in channel use $i+2$ (see Fig.~\ref{EG_F}). A similar procedure is done at the second source. 

Note that this scheme incurs a delay of 2 channel uses. Each source sends $N$ F-signals from the first channel use till channel use $i=N$. The relay feeds these signals back in the channel uses $i=2,\dots,N+1$. Finally, nodes 1 and 2 send the F-signals to their respective destinations in the channel uses $i=3,\dots,N+2$. If the F-signals $u_{j,f}$ are vectors of length $R_f$, then each source can successfully deliver $NR_f$ bits in $N+2$ channel uses. Thus the rate that each source can achieve per channel use approaches $R_f$ for large $N$.

An illustrative example for symmetric F strategy is shown in Fig.~\ref{EG_F}. The sources send $u_{j,f}(i)$ to the relay in the $i$-th channel use, which decodes the sum $u_{1,f}(i)\oplus u_{1,f}(i)$. In the same channel use, the relay feeds the signal $u_{1,f}(i-1)\oplus u_{1,f}(i-1)$ (decoded in channel use $i-1$) back to the sources. Nodes~1 and 2 use this sum to extract $u_{2,f}(i-1)$ and $u_{1,f}(i-1)$, respectively. Nodes~1 and 2 also send $u_{2,f}(i-2)$ and $u_{1,f}(i-2)$ (decoded in channel use $i-1$), respectively, to their respective destinations via the cross link in the $i$-th channel use. Nodes~3 and 4 decode $u_{1,f}(i-2)$ and $u_{2,f}(i-2)$, respectively, in channel use $i$. Note that nodes~1 and 2 always send information to the relay which renders some levels at the sources always occupied. Thus, the sources have to use {\em other levels} for sending the F-signals to the respective destinations. In general, for each F-signal, the symmetric F strategy uses 2 levels at the sources and 1 level for feedback.

Notice from Fig. \ref{EG_F} that we have sent the F-signals on levels that could have also been used by the relay to send the same amount of bits (using CN or CF). As we shall see, this symmetric F strategy does not increase the capacity if $n_c\leq n_r$. The F strategy would increase the capacity if $n_c$ is larger than $n_r$, in which case the sources would send the F-signals to their respective destinations over levels that are not accessible by the relay, thus not disturbing the relay transmission while doing so.

\subsubsection{Asymmetric} The symmetric F strategy achieves symmetric rates for the F-signals, i.e., the rate achieved by source 1 is equal to that of source 2. We can also use the F strategy in an asymmetric fashion as follows. Node 1 sends $u_{1,f}(i)$ to the relay in the $i$-th channel use, the relay decodes this signal and feeds it back to node 2 in channel use $i+1$, which sends it to node~3/destination~1 in the channel use $i+2$ {\em on the same level used by node 1}. This causes the signals $u_{1,f}(i)$ and $u_{1,f}(i-2)$ to interfere at the relay. However, the relay can always resolve this interference since it decoded $u_{1,f}(i-2)$ in channel use $i-2$. If the vector $u_{1,f}(i)$ has length $R_f$, then this strategy achieves the rate point $(R_f,0)$.

\begin{figure}[ht]
\centering
\includegraphics[width=.5\textwidth]{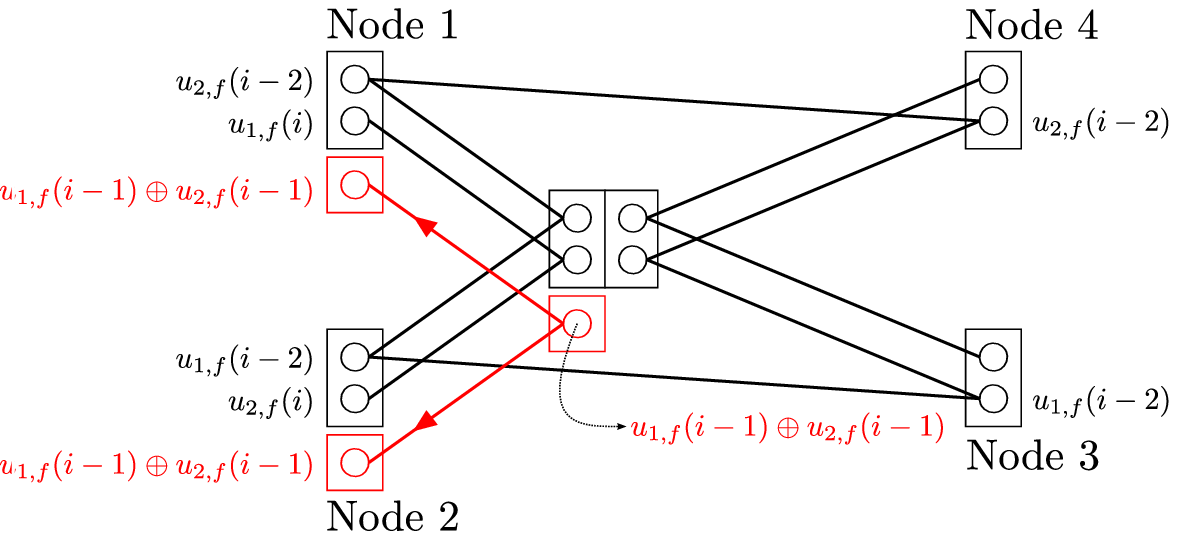}
\caption{A symmetric feedback strategy. Node~1 sends its own F-signal $u_{1,f}(i)$ to the relay to be fed back to node~2 in the next channel use. At the same time, node~1 sends node~2's F-signal $u_{2,f}(i-2)$, acquired via feedback, to node~4/destination~2. Node~2 performs similar operations. Notice the bi-directional relay channel formed by nodes~1 and 2 and the relay.}
\label{EG_F}
\vspace*{1cm}
\includegraphics[width=.5\textwidth]{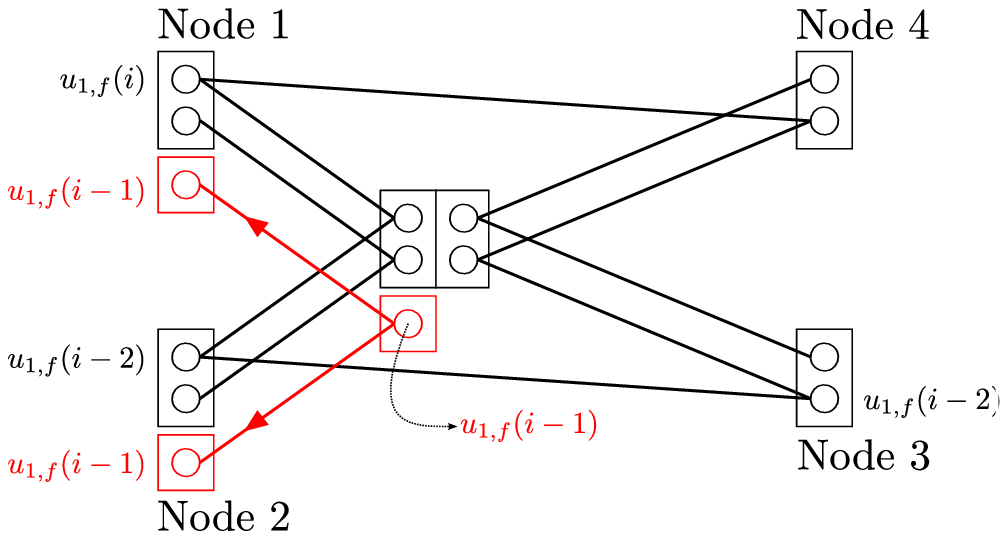}
\caption{An asymmetric feedback strategy. Node~1 sends the F-signal $u_{1,f}(i)$ to the relay. The relay decodes this signal and feeds it back to node~2 in the next channel use. Node~2 in its turn sends node~1's F-signal $u_{1,f}(i-2)$, acquired via feedback, to node~3/destination~1.}
\label{EG_F2}
\end{figure}

An illustrative example for the asymmetric F strategy is given in Fig.~\ref{EG_F2} where source~1 can send 1 bit per channel use to destination~1, achieving the rate pair $(1,0)$. Note that the same rate pair can be achieved using the symmetric F strategy (Fig. \ref{EG_F}) by setting $u_{2,f}=0$. But this would be inefficient since it consumes 2 levels at the relay for reception. The same rate pair can be achieved using the asymmetric F strategy while using only 1 level at the relay as shown in Fig. \ref{EG_F2}. This leaves one level at the relay unused, providing more flexibility to combine the F strategy with other strategies. Since our aim is to characterize the capacity region of the linear deterministic BFN with feedback, we are going to need strategies which achieve asymmetric rates efficiently. Both the symmetric and the asymmetric F strategies will be used in the sequel.

\subsection{Decode-forward}
\label{Section:CodingStrategies:DF}
\begin{figure}[ht]
\centering
\includegraphics[width=.5\textwidth]{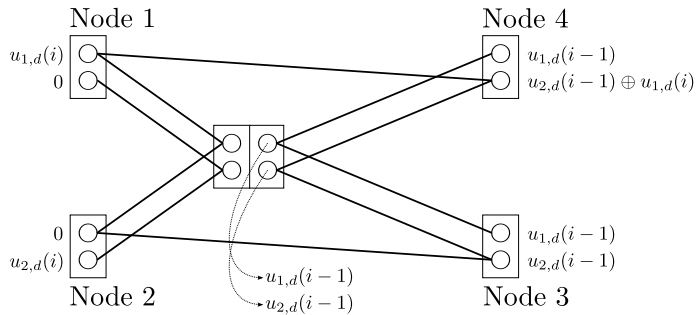}
\caption{A graphical illustration of the decode-forward strategy. Node~4/destination~2 starts be removing $u_{1,d}(i)$ (known from the decoding process in channel use $i+1$) from its received signal. Then it decodes the D-signal with time index $i-1$. Using this strategy in this setup, each source can send 1 bit per channel use.}
\label{EG_DF}
\end{figure}

The last strategy we describe in this section is the decode-forward (DF). Although this strategy is well known\cite{CoverElgamal}, we describe it here to draw the reader's attention to a convention we will adopt in the following.
In classical DF, each source sends a public signal $u_{j,d}(i)$ in the $i$-th channel use, $i=1,\dots,N$ and where the subscript $d$ is used to denote D-signals, the relay decodes both $u_{1,d}(i)$ and $u_{2,d}(i)$ in the $i$-th channel use, maps them to $u_{r,d}(i)$ which it forwards in channel use $i+1$ (see Fig.~\ref{EG_DF}). 

\begin{figure}[ht]
\centering
\includegraphics[width=.4\textwidth]{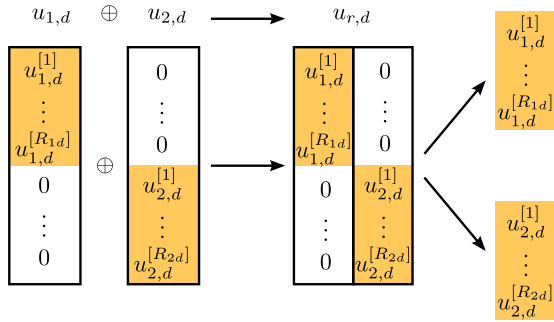}
\caption{A graphical illustration of the structure of the D-signals. Notice how $u_{1,d}$ and $u_{2,d}$ are zero-padded. Notice also that decoding the sum $u_{1,d}\oplus u_{2,d}$ is equivalent to decoding the D-signals separately. In the sequel, we will use these colored bars to represent the D-signals.}
\label{Fig:DF_Illustration}
\end{figure}

For convenience, this operation is represented as follows (see Fig. \ref{Fig:DF_Illustration}). Let the D-signal of source 1 in channel use $i$, $u_{1,d}(i)$, be a vector of length $R_{1d}+R_{2d}$ where the lower-most $R_{2d}$ positions of $u_{1,d}(i)$ are zeros. Similarly, let $u_{2,d}(i)$ be of length $R_{1d}+R_{2d}$ with zeros in the top-most $R_{1d}$ positions. Then, source 1 sends $u_{1,d}(i)$ and source 2 sends $u_{2,d}(i)$. The relay then decodes $u_{1,d}(i)\oplus u_{2,d}(i)$, a process which is equivalent to decoding both $u_{1,d}(i)$ and $u_{2,d}(i)$ separately due to the zero padding. The relay then forwards $u_{r,d}(i)=u_{1,d}(i)\oplus u_{2,d}(i)$ in channel use $i+1$ (see Fig.~\ref{EG_DF}).

The destinations start decoding from channel use $N+1$ where only the relay is active, and they both decode $u_{r,d}(N)$, which allows them to obtain both $u_{1,d}(N)$ and $u_{2,d}(N)$. Decoding proceeds backward to the $N$-th channel use. In the $N$-th channel use, the destinations start by removing $u_{j,d}(N)$ from the received signal (which they know from channel use $N+1$). Then, they decode $u_{r,d}(N-1)$ to obtain $u_{1,d}(N-1)$ and $u_{2,d}(N-1)$. In this way, the destinations obtain their desired D-signals, $R_{1d}$ bits from source 1 and $R_{2f}$ bits from source 2. Decoding proceeds backwards till the first channel use is reached. Thus, source 1 and source 2 achieve $R_{1d}$ and $R_{2d}$ bits per channel use, respectively, for large $N$. Notice that the D-signals are public since they are decoded at both destinations.

\subsection{Remark on the use of the different strategies}
At this point, a remark about the DF strategy as compared to the CN strategy in Sect. \ref{Section:CodingStrategies:CN} is in order. Due to backward decoding, the interference caused by the D-signal, $u_{1,d}(i)$ at node~4/destination~2 for instance, is not harmful since it can be removed as long as the decoding of the D-signals was successful in channel use $i+1$. This is the reason why the relay and the sources can send over the same levels at the destinations (as in Fig. \ref{EG_DF}), in contrast to CN, CF, and F where separate levels have to be allocated to the source and the relay signals. We summarize this point by saying that \emph{the D-signals $u_{j,d}(i)$ (from the sources) should be received `clean' at the relay but not necessarily so at the destinations}. In fact, the D-signals arriving from the sources do not have to be received at all at the destinations since they are decoded from the relay signal.

Now consider the N-signals where an opposite statement holds. Since the relay decodes in a forward fashion, and since the sources send `present' and `future' N-signals, i.e., $u_{j,n}(i)$ and $u_{j,n}(i+1)$ in the $i$-th channel use, then interference from the $u_{j,n}(i)$ is not harmful at the relay. This is true since this interference is known from the decoding in channel use $i-1$ at the relay, and hence can be removed. The `present' N-signal is however important at the destinations, since it is the signal that participates in interference neutralization. We summarize this statement by saying that \emph{the `present' N-signal must be received `clean' at the destinations but not necessarily so at the relay}. Additionally, \emph{the `future' N-signal must be received `clean' at the relay, but does not have to be received at all at the destinations}.

\begin{figure}[ht]
\centering
\includegraphics[width=.55\textwidth]{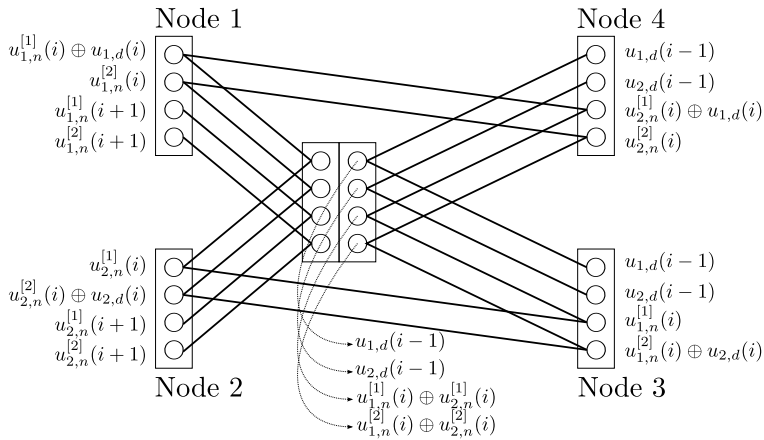}
\caption{A graphical illustration of the combination of DF and CN. The relay can obtain $u_{1,d}(i)$ and $u_{2,d}(i)$ in the $i$-th channel use after removing $u_{1,n}^{[1]}(i)\oplus u_{2,n}^{[1]}(i)$ and $u_{1,n}^{[2]}(i)\oplus u_{2,n}^{[2]}(i)$ which it has decoded in channel use $i-1$. Thus, this interference between the N-signal and the D-signals at the relay is not harmful. In the $i$-th channel use, node~3/destination~1 starts by removing $u_{2,d}(i)$ (known from the decoding process in channel use $i+1$) from its received signal. Then it decodes $u_{1,d}(i-1)$, $u_{2,d}(i-1)$, $u_{1,n}^{[1]}(i)$, and $u_{1,n}^{[2]}(i)$. Using this strategy each source can send 3 bit per channel use which achieves the sum-capacity upper bound (cf. Thm. \ref{Theorem:Capacity}).}
\label{EG_DF_CN}
\end{figure}

Combining these properties, we can construct a hybrid scheme where both CN and DF are used, and where the N-signals and the D-signals overlap at the relay and the destinations in a not harmful way,  as illustrated in Fig.~\ref{EG_DF_CN}. Here, node~1 allows its present N-signal $u_{1,n}(i)=[u_{1,n}^{[1]T}(i),\ u_{1,n}^{[2]T}(i)]^T$ to overlap with the D-signal $u_{1,d}(i)$. And thus these signals also overlap at the destination nodes. Nevertheless, the relay is still able to decode the necessary information and forward it to the destinations which can still recover their desired information. This overlap allows a more efficient exploitation of the channel levels.

In the following sections, we develop capacity achieving schemes for the linear deterministic BFN with feedback which are based on combinations of the four strategies explained above.

\section{Achievability for $n_c\leq n_r$}
\label{LowerBoundNcNr}
In this section, we show that the outer bound region given in Thm.~\ref{Theorem:CapacityOuter} is achievable for the case $n_c\leq n_r$. First, we notice that if $n_c\leq n_r$, then the feedback channel $n_f$ does not have a contribution to the upper bounds in Thm.~\ref{Theorem:CapacityOuter}, which reduces to
\begin{align*}
0\leq R_1&\leq \min\{n_s,n_r\}\\
0\leq R_2&\leq \min\{n_s,n_r\}\\
R_1+R_2&\leq n_r+n_c\\
R_1+R_2&\leq n_{r}+(n_{s}-n_{c})^+\\
R_1+R_2&\leq n_{s}+n_{c}.
\end{align*}
In this case feedback does not increase the capacity of the BFN with respect to the non-feedback case. The outer bound can be achieved without exploiting the feedback link $n_f$, and thus without using the F strategy, as per the discussion  at the end of Sect.~\ref{Section:UpperBoundsfortheld-BFNF}. Hence, in this section we only use the strategies that do not exploit feedback, i.e., CF, CN, and DF.

\subsection{Case $n_s\leq \min\{n_c,n_r\}=n_c\leq n_r$:}
\begin{lemma}
\label{Lemma:Ns_Nc_Nr}
In the linear deterministic BFN with feedback with $n_s\leq n_c\leq n_r$ the following region is achievable
\begin{align*}
0\leq R_1&\leq n_s\\
0\leq R_2&\leq n_s\\
R_1+R_2&\leq n_r,
\end{align*}
\end{lemma}
This achievable rate region coincides with the outer bound given in Thm.~\ref{Theorem:CapacityOuter}. Thus, the achievability of this region characterizes the capacity region of the linear deterministic BFN with feedback with $n_s\leq n_c\leq n_r$.

The rest of this subsection is devoted for the proof of this Lemma. In this case $\max\{n_c,n_r\}\geq n_s$ and thus we use the CF strategy according to the discussion in Sect. \ref{Section:CodingStrategies:CF}. We also use DF for achieving asymmetric rate tuples. Moreover, since $n_s\leq n_c$ we do not use the CN strategy following the discussion in Sect. \ref{Section:CodingStrategies:CN}.

\subsubsection{Encoding}
Let us construct $x_1(i)$ in the $i$-th channel use as follows
\begin{align*}
x_1(i)=\left[\begin{array}{c} 
u_{1,c}(i)\\
u_{1,d}(i)\\
0_{n_c-R_{c}-R_{1d}-R_{2d}}\\
0_{q-n_c}
\end{array}\right].
\end{align*}

The signal $u_{1,d}$ is a vector of length $R_{1d}+R_{2d}$ with the lower $R_{2d}$ components equal to zero as described in Sect. \ref{Section:CodingStrategies:DF}. Thus, it contains $R_{1d}$ information bits. The signal $u_{1,c}$ is a vector of length $R_{c}$. We construct $x_2(i)$ similarly, with $u_{2,d}$ and $u_{2,c}$ being  $(R_{1d}+R_{2d})\times1$ and $R_{c}\times1$ binary vectors, respectively, where the first $R_{1d}$ components of $u_{2,d}$ are zeros. The rates of the C-signal of both sources are chosen to be equal.

\subsubsection{Relay processing}
In the $i$-th channel use, the relay observes the top-most $n_s$ bits of $x_1(i)\oplus x_2(i)$. Under the following condition
\begin{align}
\label{Ns_Nc_Nr_RC1}
R_{c}+R_{1d}+R_{2d}\leq n_s,
\end{align}
the relay is able to observe both $u_{1,c}(i)\oplus u_{2,c}(i)$ and $u_{1,d}(i)\oplus u_{2,d}(i)$ and hence to decode them. 

\begin{figure}[ht]
\centering
\includegraphics[width=.7\columnwidth]{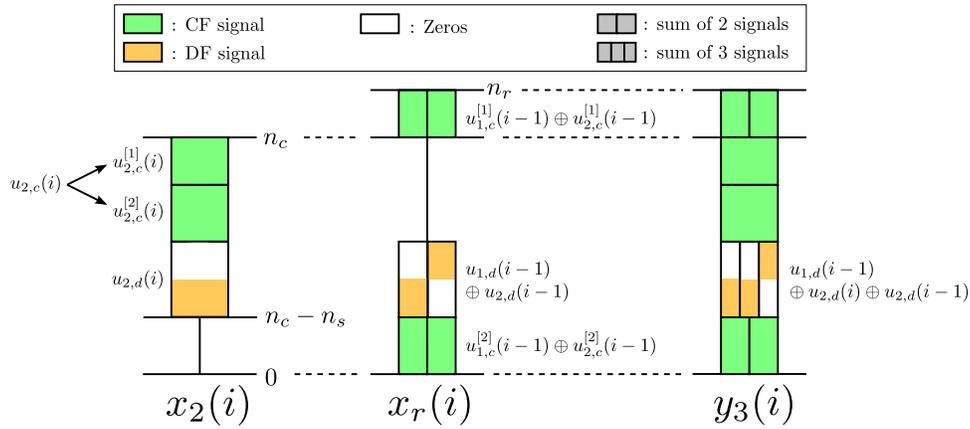}
\caption{A graphical illustration of the transmit signal of node 2, i.e., $x_2(i)$, the transmit signal of the relay, i.e., $x_r(i)$, and the received signal at node~3/destination~1, i.e., $y_3(i)$, using the capacity achieving scheme of the linear deterministic BFN with feedback with $n_s\leq n_c\leq n_r$. The color legend is shown on top.}
\label{Fig:Scheme0}
\end{figure}

Since in this case $n_r\geq n_c$, the relay can access levels at the destinations above those that can be accessed by the sources. Then, the signal $u_{1,c}(i)\oplus u_{2,c}(i)$ to be forwarded by the relay is split into two parts as follows 
\[
u_{1,c}(i)\oplus u_{2,c}(i)
=\left[\begin{array}{c}
u_{1,c}^{[1]}(i)\oplus u_{2,c}^{[1]}(i)\\
u_{1,c}^{[2]}(i)\oplus u_{2,c}^{[2]}(i)
\end{array}\right],
\]
where the upper part of length $R_c^{[1]}$ is sent such that it arrives on top of the signals from the sources, and the lower part of length $R_c^{[2]}$ is sent below, with $R_c=R_c^{[1]}+R_c^{[2]}$. Fig. \ref{Fig:Scheme0} shows the transmit signal of node~2 ($x_2(i)$) and the relay ($x_r(i)$) and received signal of node~3 ($y_3(i)$). For clarity, from this point on we drop the labels of signals that do not undergo any change from left to right in this type of pictorial illustration. Thus, the relay forwards $x_r(i+1)$ in the next channel use where
\begin{align*}
x_r(i+1)=\left[\begin{array}{c} 
0_{n_r-n_c-R_c^{[1]}}\\
u_{1,c}^{[1]}(i)\oplus u_{2,c}^{[1]}(i)\\
0_{R_{c}}\\
u_{1,d}(i)\oplus u_{2,d}(i)\\
u_{1,c}^{[2]}(i)\oplus u_{2,c}^{[2]}(i)\\
0_{n_c-R_c^{[1]}-2R_{c}^{[2]}-R_{1d}-R_{2d}}\\
0_{q-n_r}
\end{array}\right].
\end{align*}
This construction requires
\begin{align}
\label{Ns_Nc_Nr_RC2a}
R_c^{[1]}+2R_{c}^{[2]}+R_{1d}+R_{2d}&\leq n_c,\\
\label{Ns_Nc_Nr_RC2b}
R_{c}^{[1]}&\leq n_r-n_c.
\end{align}

\subsubsection{Decoding at the destinations}
Node~3/destination~1 waits until the end of channel use $N+1$ where only the relay is active, and it receives the top-most $n_r$ bits of $x_r(N+1)$. Then, it decodes $u_{1,c}^{[1]}(N)\oplus u_{2,c}^{[1]}(N)$, $u_{1,c}^{[2]}(N)\oplus u_{2,c}^{[2]}(N)$, and $u_{1,d}(N)\oplus u_{2,d}(N)$. Similarly at the second receiver. At this point, both receivers have obtained both D-signals $u_{1,d}(N)$ and $u_{2,d}(N)$ which they extract from $u_{1,d}(N)\oplus u_{2,d}(N)$ (recall our discussion in Sect.~\ref{Section:CodingStrategies:DF}).

\begin{figure}[ht]
\centering
\includegraphics[width=.7\columnwidth]{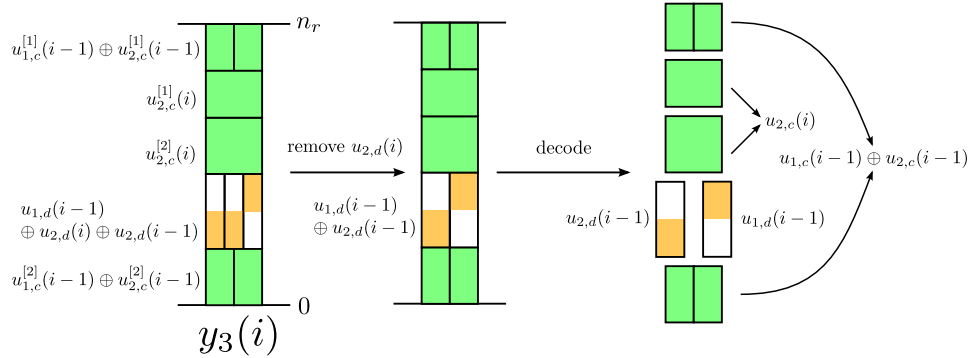}
\caption{The decoding steps at the destination. Due to backward decoding, node~3/destination~1 knows $u_{2,d}(i)$ and $u_{1,c}(i)\oplus u_{2,c}(i)$ when decoding $y_3(i)$ (decoded in channel use $i+1$). It starts by removing $u_{2,d}(i)$ from $y_3(i)$. Then it decodes the C-signal sum $u_{1,c}(i-1)\oplus u_{2,c}(i-1)$, the C-signal interference $u_{2,c}(i)$, and the D-signals $u_{1,d}(i-1)$ and $u_{2,d}(i-1)$. Finally, it uses the CF sum $u_{1,c}(i)\oplus u_{2,c}(i)$ and $u_{2,c}(i)$ to extract $u_{1,c}(i)$.}
\label{Fig:Scheme0_Decoding}
\end{figure}

The destinations proceed to the $N$-th channel use. The received signal at node~3/destination~1 can be written as (see Fig.~\ref{Fig:Scheme0} or \ref{Fig:Scheme0_Decoding})
\begin{align*}
y_3(N)=\left[\begin{array}{c} 
0_{q-n_r}\\
0_{n_r-n_c-R_c^{[1]}}\\
u_{1,c}^{[1]}(N-1)\oplus u_{2,c}^{[1]}(N-1)\\
u_{2,c}(N)\\
u_{2,d}(N)\oplus u_{1,d}(N-1)\oplus u_{2,d}(N-1)\\
u_{1,c}^{[2]}(N-1)\oplus u_{2,c}^{[2]}(N-1)\\
0_{n_c-R_c^{[1]}-2R_{c}^{[2]}-R_{1d}-R_{2d}}
\end{array}\right].
\end{align*}
Since node~3/destination~1 knows $u_{2,d}(N)$, it can remove it from the received signal (see Fig. \ref{Fig:Scheme0_Decoding}). Then it proceeds with decoding 
$$u_{1,c}^{[1]}(N-1)\oplus u_{2,c}^{[1]}(N-1),\quad u_{1,c}^{[2]}(N-1)\oplus u_{2,c}^{[2]}(N-1),\quad u_{2,c}(N),$$
$$\text{and }\quad u_{1,d}(N-1)\oplus u_{2,d}(N-1)$$
Having $u_{2,c}(N)$ allows node~3/destination~1 to obtain $u_{1,c}(N)$ as $u_{1,c}(N)\oplus u_{2,c}(N)\oplus u_{2,c}(N)=u_{1,c}(N)$. Additionally, node~3/destination~1 obtains $u_{1,d}(N-1)$ which is a desired signals. Furthermore, $u_{2,d}(N-1)$ and $u_{1,c}(N-1)\oplus u_{2,c}(N-1)$ are obtained which are used in the decoding process in channel use $N-1$.

In this process, node~3/destination~1 was able to recover its C and its D-signals comprising of $R_{c}$ and $R_{1d}$ bits, respectively. Node~4 performs similar operations. The receivers proceed backwards till channel use 1 is reached.

\subsubsection{Achievable region}
The rates achieved by source~1 and 2 are $R_1=R_{c}+R_{1d}$ and $R_2=R_{c}+R_{2d}$, respectively. Collecting the rate constraints \eqref{Ns_Nc_Nr_RC1}, \eqref{Ns_Nc_Nr_RC2a}, and \eqref{Ns_Nc_Nr_RC2b}, we get the following constraints on the non-negative rates $R_{c}^{[1]}$, $R_{c}^{[2]}$, $R_{1d}$, and $R_{2d}$:
\begin{align*}
R_{c}^{[1]}+R_c^{[2]}+R_{1d}+R_{2d}&\leq n_s\\
R_c^{[1]}+2R_{c}^{[2]}+R_{1d}+R_{2d}&\leq n_c\\
R_{c}^{[1]}&\leq n_r-n_c.
\end{align*}
Using Fourier-Motzkin's elimination \cite[Appendix D]{ElgamalKim} we get the following achievable region
\begin{align*}
0\leq R_1&\leq n_s\\
0\leq R_2&\leq n_s\\
R_1+R_2&\leq n_r,
\end{align*}
which proves Lemma \ref{Lemma:Ns_Nc_Nr}.

\subsection{Case $n_s>\min\{n_c,n_r\}$ or equivalently $n_c\leq\min\{n_s,n_r\}$:}
Now we consider the case where $n_c\leq\min\{n_s,n_r\}$, i.e., the cross channel is weaker than both the source-relay channel and the relay-destination channel. As we have mentioned earlier, if $n_s\geq n_c$, then we can pass some future information to the relay without the destinations noticing by using the CN strategy. Thus, we use CN in addition to CF and DF. For this case, we have the following lemma.
\begin{lemma}
\label{Lemma:Nc_NsNr}
The rate region defined by the following rate constraints
\begin{align*}
0\leq R_1&\leq \min\{n_s,n_r\}\\
0\leq R_2&\leq \min\{n_s,n_r\}\\
R_1+R_2&\leq n_s+n_c\\
R_1+R_2&\leq n_r+n_c\\
R_1+R_2&\leq n_r+n_s-n_c,
\end{align*}
is achievable in the linear deterministic BFN with feedback with $n_c\leq\min\{n_s,n_r\}$.
\end{lemma} 
This region coincides with the outer bound given in Thm.~\ref{Theorem:CapacityOuter}. Thus, the scheme which achieves this region achieves the capacity of the linear deterministic BFN with feedback with $n_c\leq\min\{n_s,n_r\}$. We provide this capacity achieving scheme in the rest of this subsection.

\subsubsection{Encoding}
At time instant $i$, node 1 sends the following signal
\begin{align*}
x_1(i)=\left[\begin{array}{c}
0_{n_c-R_c-R_{1d}-R_{2d}-R_n}\\
u_{1,c}(i)\\
u_{1,d}(i)\\
\left[\begin{array}{c}
u_{1,n}(i)\\
0_{n_s-n_c-R_n}\end{array}\right]\oplus\left[\begin{array}{c}
0_{n_s-n_c-\overline{R}_{1d}-\overline{R}_{2d}}\\
\overline{u}_{1,d}(i)\end{array}\right]\\
u_{1,n}(i+1)\\
0_{q-n_s}
\end{array}\right],
\end{align*}
where $u_{1,n}(i+1)$ is the future information passed to the relay. The signals $u_{1,c}$, $u_{1,d}$, $u_{1,n}$, and $\overline{u}_{1,d}$ are vectors of length $R_c$, $R_{1d}+R_{2d}$, $R_n$, and $\overline{R}_{1d}+\overline{R}_{2d}$, respectively. Notice that this construction requires that
\begin{align}
\label{RC1}
R_c+R_{1d}+R_{2d}+R_{n}&\leq n_c\\
\overline{R}_{1d}+\overline{R}_{2d}&\leq n_s-n_c\\
\label{RC3}
R_{n}&\leq n_s-n_c.
\end{align}

\begin{figure}[ht]
\centering
\includegraphics[width=.65\columnwidth]{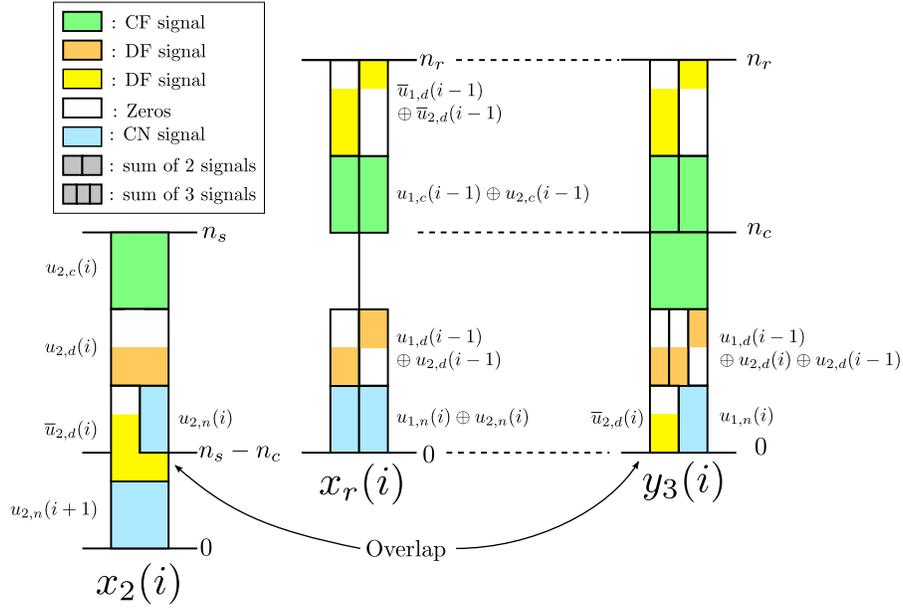}
\caption{A graphical illustration of the $x_2(i)$, $x_r(i)$, and $y_3(i)$ for the capacity achieving scheme of the linear deterministic BFN with feedback with $n_c\leq \min\{n_s,n_r\}$. Notice the overlap of $\overline{u}_{j,d}$ and $u_{j,n}$.}
\label{Fig:Scheme1}
\end{figure}

Using this construction, there can be an overlap between $u_{1,n}(i)$ and $\overline{u}_{1,d}(i)$ at the relay and at the destinations (see Fig. \ref{Fig:Scheme1}). However, this overlap is not harmful (similar to the one discussed in Sect. \ref{Section:CodingStrategies:DF}). The overlapping D-signal is marked with an overline to distinguish it from $u_{1,d}$ which does not overlap with any signal at the relay.

A similar construction is employed by the second source. As we show next, this construction allows us to achieve the capacity region of the linear deterministic BFN with feedback in this case. The task now is to find the conditions that $R_{1d}$, $R_{2d}$, $\overline{R}_{1d}$. $\overline{R}_{2d}$ $R_{c}$, and $R_{n}$ should satisfy in order to guarantee reliable decoding.

\subsubsection{Relay Processing}
The received signal at the relay consists of the top $n_{s}$ bits of $x_1(i)\oplus x_2(i)$.
Let us write $y_0(i)$ as follows
\begin{align*}
y_0(i)=\left[\begin{array}{c}
0_{q-n_s}\\
0_{n_c-R_c-R_{1d}-R_{2d}-R_n}\\
u_{1,c}(i)\oplus u_{2,c}(i)\\
u_{1,d}(i)\oplus u_{2,d}(i)\\
\left[\begin{array}{c}
u_{1,n}(i)\oplus u_{2,n}(i)\\
0_{n_s-n_c-R_n}\end{array}\right]\oplus \left[\begin{array}{c}
0_{n_s-n_c-\overline{R}_{1d}-\overline{R}_{2d}}\\
\overline{u}_{1,d}(i)\oplus \overline{u}_{2,d}(i)\end{array}\right]\\
u_{1,n}(i+1)\oplus u_{2,n}(i+1)\\
\end{array}\right].
\end{align*}

\begin{figure}[ht]
\centering
\includegraphics[width=.55\columnwidth]{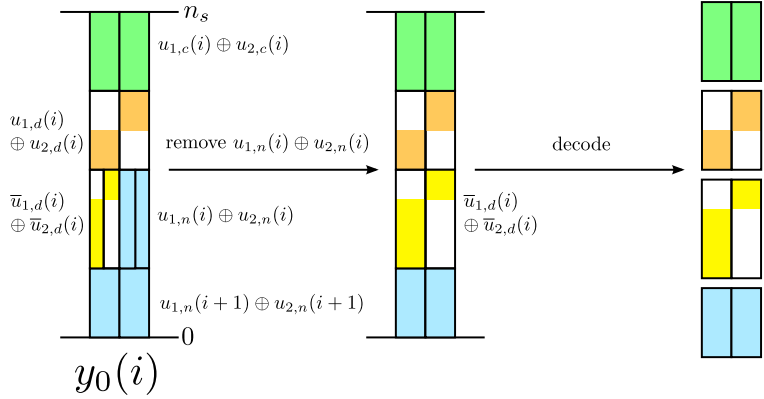}
\caption{The relay receives the superposition of $x_1(i)$ and $x_2(i)$ in the $i$-th channel use. First, it removes $u_{2,n}(i)\oplus u_{1,n}(i)$ which it knows from the decoding process in channel use $i-1$. Next, it decodes $u_{1,n}(i+1)\oplus u_{2,n}(i+1)$, $u_{1,c}(i)\oplus u_{2,c}(i)$, $u_{1,d}(i)\oplus u_{2,d}(i)$, and $\overline{u}_{1,d}(i)\oplus\overline{u}_{2,d}(i)$ which it forwards in channel use $i+1$ as shown in Fig. \ref{Fig:Scheme1}.}
\label{Fig:Scheme1_Relay_Decoding}
\end{figure}

In the $i$-th channel use, the relay knows  $u_{1,n}(i)\oplus u_{2,n}(i)$ from the decoding process in the channel use $i-1$. This allows it to remove $u_{1,n}(i)\oplus u_{2,n}(i)$ from $y_0(i)$ (see Fig. \ref{Fig:Scheme1_Relay_Decoding}). Then, the relay can decode $u_{1,c}(i)\oplus u_{2,c}(i)$, $u_{1,d}(i)\oplus u_{2,d}(i)$, $\overline{u}_{1,d}(i)\oplus \overline{u}_{2,d}(i)$, and finally $u_{1,n}(i+1)\oplus u_{2,n}(i+1)$. At the end of channel use $i$, the relay constructs the following signal
\begin{align}
\label{Xri+1}
x_r(i+1)=\left[\begin{array}{c} 
0_{n_r-\overline{R}_{1d}-\overline{R}_{2d}-2R_c-R_{1d}-R_{2d}-R_n}\\
\overline{u}_{1,d}(i)\oplus \overline{u}_{2,d}(i)\\
u_{1,c}(i)\oplus u_{2,c}(i)\\
0_{R_{c}}\\
u_{1,d}(i)\oplus u_{2,d}(i)\\
u_{1,n}(i+1)\oplus u_{2,n}(i+1)\\
0_{q-n_r}
\end{array}\right]
\end{align}
and sends it in channel use $i+1$. The constituent signals of $x_r(i+1)$ in \eqref{Xri+1} fit in an interval of size $n_r$ if
\begin{align}
\label{RC4}
\overline{R}_{1d}+\overline{R}_{2d}+2R_c+R_{1d}+R_{2d}+R_n\leq n_r.
\end{align}

\subsubsection{Decoding at the destinations}
In the following, consider the processing at node~3/destination~1 (the processing at node~4/destination~2 follows similar lines). Node~3/destination~1 will observe the top-most $n_c$ bits of $x_2(i)$ plus the top-most $n_r$ bits of $x_r(i)$ (modulo-2) at the $i$-th channel use. That is, we can write the received signal at node~3/destination~1 as
\begin{align*}
y_3(i)=\left[\begin{array}{c} 
0_{q-n_r}\\
0_{n_r-\overline{R}_{1d}-\overline{R}_{2d}-2R_c-R_{1d}-R_{2d}-R_n}\\
\overline{u}_{1,d}(i-1)\oplus \overline{u}_{2,d}(i-1)\\
u_{1,c}(i-1)\oplus u_{2,c}(i-1)\\
u_{2,c}(i)\\
u_{1,d}(i-1)\oplus u_{2,d}(i-1)\oplus u_{2,d}(i)\\
u_{1,n}(i)\oplus \overline{u}_{2,d}^u(i)
\end{array}\right]
\end{align*}
where we used $\overline{u}^u_{2,d}(i)$ to denote the top-most $(n_c-n_s+\overline{R}_{1d}+\overline{R}_{2d}+R_{n})^+$ bits of $\overline{u}_{2,d}(i)$. Notice the effect of CN: node~3/destination~1 receives $u_{1,n}(i)$ interference free (except the interference caused by the previously decoded $\overline{u}^u_{2,d}(i)$ which is not harmful) on the lowest $R_n$ levels.

\begin{figure}[ht]
\centering
\includegraphics[width=.65\columnwidth]{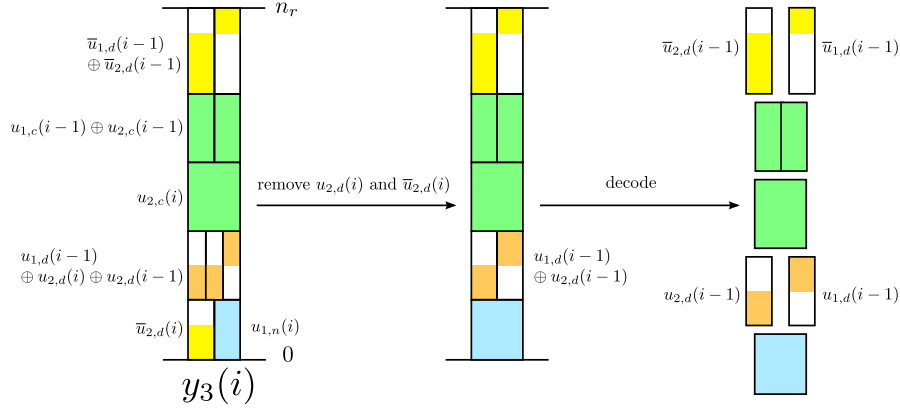}
\caption{The decoding process at node~3/destination~1. First, the receiver removes $u_{2,d}(i)$ and $\overline{u}_{2,d}(i)$ from $y_3(i)$ which it knows from the decoding process in channel use $i+1$ (backward decoding). Then, it decodes the signals $\overline{u}_{1,d}(i-1)\oplus \overline{u}_{2,d}(i-1)$, $u_{1,c}(i-1)\oplus u_{2,c}(i-1)$, followed by $u_{2,c}(i)$, $u_{1,d}(i-1)\oplus u_{2,d}(i-1)$ and $u_{1,n}(i)$. Notice how cooperative neutralization CN allows node~3/destination~1 to decode $u_{1,n}(i)$ interference free. Finally, $u_{1,c}(i)$ is extracted from $u_{2,c}(i)$ and $u_{1,c}(i)\oplus u_{2,c}(i)$ (known from the decoding process in block $i+1$).}
\label{Fig:Scheme1_Decoding}
\end{figure}

Decoding at the receivers proceeds backwards. As shown in Fig. \ref{Fig:Scheme1_Decoding}, node~3/destination~1 starts with removing $\overline{u}_{2,d}^u(i)$ and $u_{2,d}(i)$. Then, it decodes $\overline{u}_{1,d}(i-1)$, $\overline{u}_{2,d}(i-1)$, $u_{1,c}(i-1)\oplus u_{2,c}(i-1)$, $u_{2,c}(i)$, $u_{1,d}(i-1)$, $u_{2,d}(i-1)$, and $u_{1,n}(i)$. Decoding then proceeds backwards till $i=1$.

\subsubsection{Achievable region}
Collecting the rate constraints \eqref{RC1}-\eqref{RC3} and \eqref{RC4}, we conclude that the non-negative rates $\overline{R}_{1,d}$, $\overline{R}_{2,d}$, $R_{1,d}$, $R_{2,d}$, $R_{c}$, and $R_{n}$ can be achieved if they satisfy
\begin{align*}
R_c+R_{1d}+R_{2d}+R_{n}&\leq n_c\\
\overline{R}_{1d}+\overline{R}_{2d}&\leq n_s-n_c\\
R_{n}&\leq n_s-n_c\\
\overline{R}_{1d}+\overline{R}_{2d}+2R_c+R_{1d}+R_{2d}+R_n&\leq n_r.
\end{align*}
Using Fourier Motzkin's elimination with $R_1=R_{1d}+\overline{R}_{1d}+R_c+R_n$ and $R_2=R_{2d}+\overline{R}_{2d}+R_c+R_n$, we can show that the following region is achievable 
\begin{align*}
0\leq R_1&\leq \min\{n_s,n_r\}\\
0\leq R_2&\leq \min\{n_s,n_r\}\\
R_1+R_2&\leq n_s+n_c\\
R_1+R_2&\leq n_r+n_c\\
R_1+R_2&\leq n_r+n_s-n_c,
\end{align*}
which proves Lemma \ref{Lemma:Nc_NsNr}. At this point, we have finished the proof of the achievability of Thm.~\ref{Theorem:Capacity} for $n_c\leq n_r$.

\section{Achievability for $n_c>n_r$}
\label{LowerBoundNrNc}
In this section, we prove Thm.~\ref{Theorem:Capacity} for the BFN with $n_c>n_r$, which reduces to
\begin{align*}
R_1&\leq \min\{n_s,n_r+n_f,n_c\}\\
R_2&\leq \min\{n_s,n_r+n_f,n_c\}\\
R_1+R_2&\leq n_{c}+[n_{s}-n_{c}]^+.
\end{align*}
In this case, $n_f$ contributes to the outer bounds. If the region defined by these upper bounds is achievable, then feedback has a positive impact on the BFN. This is what we shall prove next. That is, we show that this region is in fact achievable, and hence that relay-source feedback increases the capacity of the network if $n_c>n_r$ when compared to the case $n_f=0$.

We first show a toy example to explain the main ingredients of the achievable scheme. Then, depending on the relation between $n_s$ and $n_c$, we split the proof of the achievability of Thm.~\ref{Theorem:CapacityOuter} for $n_c>n_r$ to two cases: $\max\{n_r,n_s\}<n_c$ and $n_r<n_c\leq n_s$.

\subsection{A toy example: feedback enlarges the capacity region.}
\begin{figure}
\centering
\includegraphics[width=.25\textwidth]{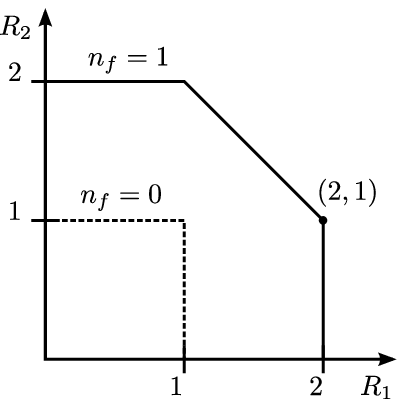}
\caption{Capacity regions for the deterministic BFN with $(n_c,n_s,n_r)=(2,3,1)$. Dotted line: without feedback ($n_f=0$); solid line: with feedback with $n_f=1$.}
\label{EG_Region}
\vspace*{1cm}
\includegraphics[width=.45\textwidth]{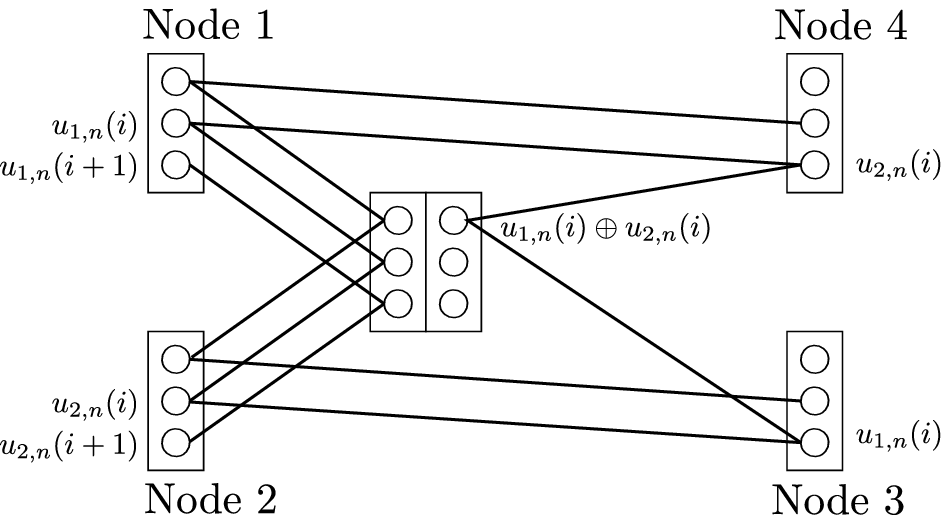}
\caption{The deterministic BFN with $(n_c,n_s,n_r)=(2,3,1)$ and $n_f=0$. The given scheme achieves the corner point $(1,1)$ of the capacity region in Fig.~\ref{EG_Region}.}
\label{TEF1}
\vspace*{1cm}
\includegraphics[width=.45\textwidth]{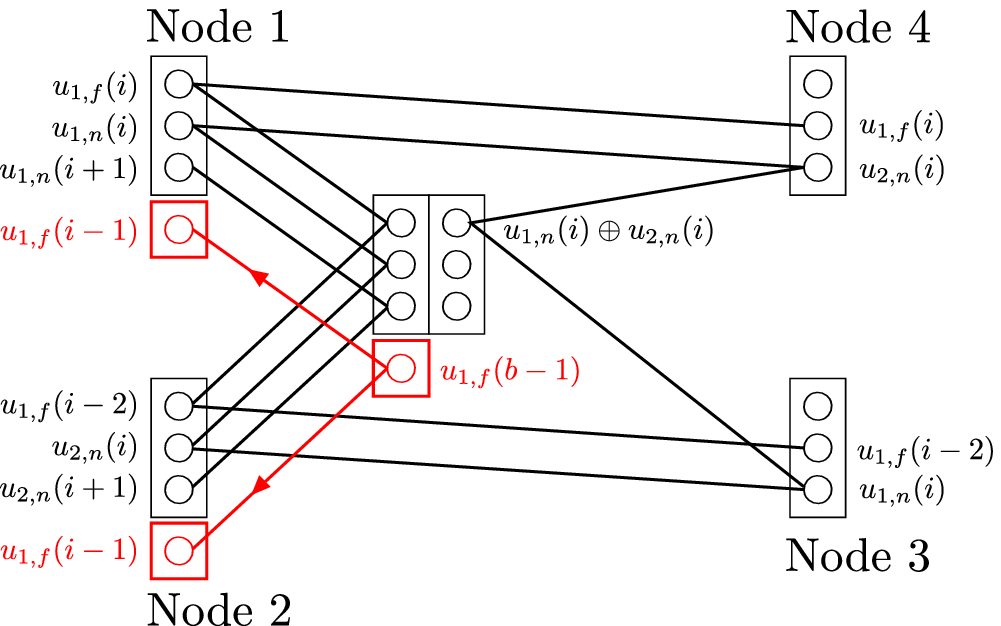}
\caption{The deterministic BFN with $(n_c,n_s,n_r)=(2,3,1)$ with $n_f=1$. The sources use the same scheme as in Fig. \ref{TEF1} to achieve 1 bit each, and source 1 uses feedback to achieve one additional bit per channel use, thus enlarging the acievable region and achieving the optimal corner point $(2,1)$ in Fig.~\ref{EG_Region}. In the $i$-th channel use, node~1 sends $u_{1,f}(i)$ to the relay, the relay feeds back $u_{1,f}(i-1)$, which it decoded in channel use $i-1$, and node~2 sends $u_{1,f}(i-2)$ to node~3/destination~1.}
\label{FH2}
\end{figure}

Consider a linear deterministic BFN with feedback where $(n_c,n_s,n_r)=(2,3,1)$. According to the upper bounds above, the capacity of this setup is outer bounded by 
\begin{align*}
R_1&\leq 1\\
R_2&\leq 1
\end{align*}
if $n_f=0$ (please refer to Fig.~\ref{EG_Region}). This capacity region is achieved using CN as shown in Fig.~\ref{TEF1}.

Now assume that this network has a feedback channel from the relay to the sources with capacity $n_f=1$. The capacity in this case is outer bounded by 
\begin{align*}
R_1&\leq 2\\
R_2&\leq 2\\
R_1+R_2&\leq 3
\end{align*}
as shown in Fig.~\ref{EG_Region}. In this case, the two sources can use the F strategy in Sect. \ref{Section:CodingStrategies:F} to exchange messages among each other. Then, the sources can use their cross channels $n_c$ to send some more bits and achieve higher rates. This idea is illustrated and described in the caption of Fig.~\ref{FH2}, where we show how to achieve the corner point $(2,1)$ in Fig.~\ref{EG_Region}. The other corner point can be achieved similarly by swapping the roles of the sources. The corner points $(2,0)$ and $(0,2)$ can be achieved by setting $u_{2,n}$ and $u_{1,n}$ to zero, respectively. The whole region is achievable by time sharing between corner points, and hence this scheme is optimal.

\subsection{Case $\max\{n_r,n_s\}<n_c$}
We start by stating the achievable region described in this subsection in the following lemma.
\begin{lemma}
\label{Lemma:NrNs_Nc}
The rate region defined by the following inequalities
\begin{align*}
0\leq R_1&\leq\min\{n_s,n_r+n_f\}\\
0\leq R_2&\leq\min\{n_s,n_r+n_f\}\\
R_1+R_2&\leq n_c,
\end{align*}
is achievable in the linear deterministic BFN with feedback with $\max\{n_r,n_s\}<n_c$.
\end{lemma}
Notice that this region matches the outer bound given in Thm.~\ref{Theorem:CapacityOuter}. Therefore, the achievability of this region proves the achievability of Thm.~\ref{Theorem:Capacity} for $\max\{n_r,n_s\}<n_c$. We prove this lemma in the rest of this subsection.

Since $n_c<n_r$ the sources can use the upper $n_c-n_r$ levels at the destinations which are not accessible by the relay to send feedback information to the destinations. Thus, in this case we use the F strategy. Since $\max\{n_c,n_r\}>n_s$ in this case, we also use the CF strategy following the intuition in Sect. \ref{Section:CodingStrategies:CF}. Furthermore, we use DF to achieve asymmetric rate pairs. The capacity achieving scheme is described next.

\subsubsection{Encoding}
\begin{figure}[ht]
\centering
\includegraphics[width=0.65\columnwidth]{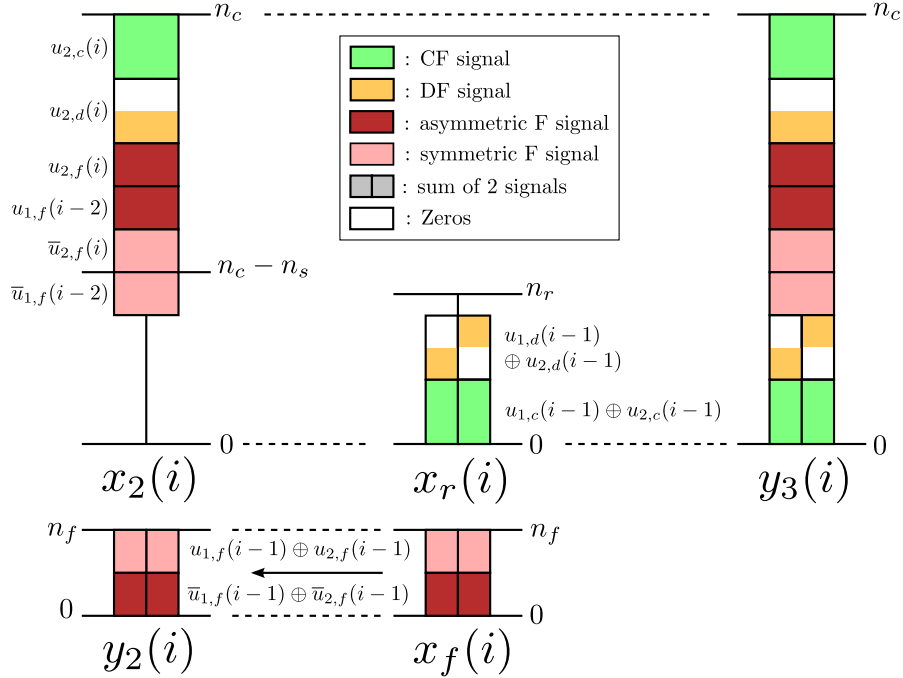}
\caption{A graphical illustration of the transmit and received signal at node~2, i.e., $x_2(i)$ and $y_2(i)$, the relay signal $x_r(i)$, the feedback signal $x_f(i)$, and the received signal at node~3/destination~1 $y_3(i)$, for the capacity achieving scheme of the linear deterministic BFN with feedback with $\max\{n_r,n_s\}\leq n_c$ with $n_f>0$. Node 2 uses the received feedback signals to extract $u_{1,f}(i-1)$ and $\overline{u}_{1,f}(i-1)$ which it sends to node~3/destination~1 in channel use $i+1$. Node~3/destination~1 uses $u_{1,c}(i)\oplus u_{2,c}(i)$ (decoded in time channel use $i+1$) and $u_{2,c}(i)$ to extract $u_{1,c}(i)$.}
\label{Fig:Scheme3}
\end{figure}

In the $i$-th channel use, node 1 sends the following signal (as shown in Fig. \ref{Fig:Scheme3})
\begin{align}
\label{NsNr_Nc_X1i}
x_1(i)=\left[\begin{array}{c} 
u_{1,c}(i)\\
u_{1,d}(i)\\
u_{1,f}(i)\\
u_{2,f}(i-2)\\
\overline{u}_{1,f}(i)\\
\overline{u}_{2,f}(i-2)\\
0_{n_c-R_{c}-R_{1d}-R_{2d}-R_{1f}-R_{2f}-2\overline{R}_f}\\
0_{q-n_{c}}
\end{array}\right].
\end{align}
Here, the signals $u_{1,f}$ is the signals used to establish the asymmetric F strategy, which is a vector of length $R_{1f}$. Similarly, $u_{2,f}(i-2)$ is the F-signal of node 2 of length $R_{2f}$, and is available at node 1 via feedback. The signal $\overline{u}_{1,f}$ is the signal used in the symmetric F strategy, and is a vector of length $\overline{R}_f$. We use both symmetric feedback and asymmetric feedback to achieve all points on the closure of the region given in Lemma \ref{Lemma:NrNs_Nc}. The C-signal $u_{1,c}$ has length $R_c$, and the D-signal $u_{1,d}$ has length $R_{1d}+R_{2d}$, containing information in the upper $R_{1d}$ bits and zeros in the lower $R_{2d}$ bits as described in Sect. \ref{Section:CodingStrategies:DF}. Node 2 sends a similar signal, with $u_{2,d}$ having zeros in the upper $R_{1d}$ positions, and with $u_{1,f}(i)$ and $u_{2,f}(i-2)$ replaced by $u_{1,f}(i-2)$ and $u_{2,f}(i)$, respectively. The given construction works if 
\begin{align}
\label{RCWFNsNc0}
R_{c}+R_{1d}+R_{2d}+R_{1f}+R_{2f}+2\overline{R}_f\leq n_c.
\end{align}

\subsubsection{Relay processing}
The relay observes the top-most $n_s$ bits of $x_1(i)\oplus x_2(i)$. We want the relay to be able to observe $u_{1,c}(i)\oplus u_{2,c}(i)$, $u_{1,d}(i)\oplus u_{2,d}(i)$, $u_{1,f}(i)\oplus u_{1,f}(i-2)$, $u_{2,f}(i-2)\oplus u_{2,f}(i)$, and $\overline{u}_{1,f}(i)\oplus \overline{u}_{2,f}(i)$. This is possible if we choose 
\begin{align}
\label{RCWFNsNc1}
R_{c}+R_{1d}+R_{2d}+R_{1f}+R_{2f}+\overline{R}_f\leq n_s.
\end{align}

\begin{figure}[ht]
\centering
\includegraphics[width=.68\columnwidth]{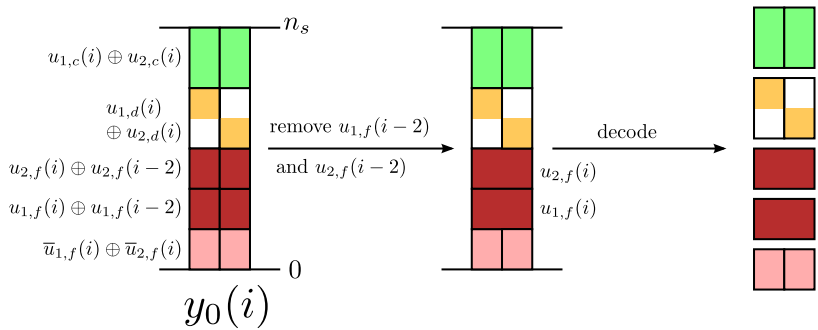}
\caption{The decoding process at the relay. In the $i$-th channel use, the relay starts with removing the known signals $u_{1,f}(i-2)$ and $u_{2,f}(i-2)$ (which it decoded in channel use $i-2$). Then, it decodes $u_{1,c}(i)\oplus u_{2,c}(i)$, $u_{1,d}(i)\oplus u_{2,d}(i)$ and $\overline{u}_{1,f}(i)\oplus \overline{u}_{2,f}(i)$ as well as $u_{2,f}(i)$ and $u_{1,f}(i)$. The F-signals are then sent back to the sources, and the DF and C-signals to the destinations as illustrated in Fig. \ref{Fig:Scheme3}.}
\label{Fig:Scheme3_Relay_Decoding}
\end{figure}

Given this condition is satisfied, the relay starts by removing $u_{1,f}(i-2)$ and $u_{2,f}(i-2)$ (known from past decoding) from $y_0(i)$ as shown in Fig. \ref{Fig:Scheme3_Relay_Decoding}. Next, it decodes the sum of the C-signals $u_{1,c}(i)\oplus u_{2,c}(i)$, the sum of the D-signals $u_{1,d}(i)\oplus u_{2,d}(i)$, the F-signals $u_{1,f}(i)$ and $u_{2,f}(i)$ and $\overline{u}_{1,f}(i)\oplus \overline{u}_{2,f}(i)$. Then it sends the following signal
\begin{align*}
x_{r}(i+1)=\left[\begin{array}{c} 
0_{n_r-R_{c}-R_{1d}-R_{2d}}\\
u_{1,d}(i)\oplus u_{2,d}(i)\\
u_{1,c}(i)\oplus u_{2,c}(i)\\
0_{q-n_r}
\end{array}\right],
\end{align*}
over the forward channel (relay-destination channel) in channel use $i+1$, which requires
\begin{align}
\label{RCWFNsNc2}
R_{c}+R_{1d}+R_{2d}\leq n_r.
\end{align}
It also sends the feedback signal $x_f(i+1)$ on the backward channel (feedback channel), where
\begin{align*}
x_{f}(i+1)=\left[\begin{array}{c} 
u_{1,f}(i)\oplus u_{2,f}(i)\\
\overline{u}_{1,f}(i)\oplus \overline{u}_{2,f}(i)\\
0_{q-R_f}
\end{array}\right],
\end{align*}
in channel use $i+1$. The signal $x_f$ represents feedback to nodes 1 and 2. Note that we feed back the signal $u_{1,f}\oplus u_{2,f}(i)$ instead of separately sending $u_{1,f}$ and $u_{2,f}(i)$. This allows a more efficient use of the feedback channel. If the vectors $u_{1,f}$ and $u_{2,f}(i)$ have different lengths, the shorter is zero padded till they have equal length.

The construction of the feedback signal $x_f(i+1)$ requires
\begin{align}
\label{RCWFNsNc3a}
R_{1f}+\overline{R}_f&\leq n_f\\
\label{RCWFNsNc3b}
R_{2f}+\overline{R}_f&\leq n_f.
\end{align}

\subsubsection{Processing feedback at the sources}
Consider node~1 at time instant $i+1$. Node 1 receives the feedback signal given by
\begin{align*}
y_1(i+1)&=\mathbf{S}^{q-n_f}x_f(i+1)=\left[\begin{array}{c} 0_{q-n_f}\\
u_{1,f}(i)\oplus u_{2,f}(i)\\
\overline{u}_{1,f}(i)\oplus \overline{u}_{2,f}(i)\\
0_{n_f-R_f}\end{array}\right].
\end{align*}
Node~1 decodes $u_{1,f}(i)\oplus u_{2,f}(i)$ and $\overline{u}_{1,f}(i)\oplus \overline{u}_{2,f}(i)$. Since node 1 knows its own F-signal $u_{1,f}(i)$, then it can extract $u_{2,f}(i)$ from this feedback information. Similarly, it can extract $\overline{u}_{2,f}(i)$. Therefore, in channel use $i+2$, node 1 knows the F-signals of node~2 which are $u_{2,f}(i)$ and $\overline{u}_{2,f}(i)$ which justifies the transmission of $u_{2,f}(i-2)$ and $\overline{u}_{2,f}(i-2)$ in $x_1(i)$ in \eqref{NsNr_Nc_X1i}. After processing this feedback, node~1 is able to send node~2's F-signals to node~4/destination~2. A similar processing is performed at node~2, which sends the F-signals of node~1 to node~3/destination~1.

\subsubsection{Decoding at the destinations}
Assume that
\begin{align}
\label{RCWFNsNc4}
2R_{c}+2R_{1d}+2R_{2d}+R_{1f}+R_{2f}+2\overline{R}_f\leq n_c.
\end{align}
In this case, node~3/destination~1 for instance is able to observe all the signals sent by node~2 and the relay. The received signal $y_3(i)$ is then
\begin{align*}
y_{3}(i)=\left[\begin{array}{c} 
0_{q-n_c}\\
u_{2,c}(i)\\
u_{2,d}(i)\\
u_{2,f}(i)\\
u_{1,f}(i-2)\\
\overline{u}_{2,f}(i)\\
\overline{u}_{1,f}(i-2)\\
0_{n_c-2R_{c}-2R_{1d}-2R_{2d}-R_{1f}-R_{2f}-2\overline{R}_f}\\
u_{1,d}(i-1)\oplus u_{2,d}(i-1)\\
u_{1,c}(i-1)\oplus u_{2,c}(i-1)\\
\end{array}\right].
\end{align*}
Node~3/destination~1 decodes backwards starting with $i=N+2$ where the desired F-signals $u_{1,f}(N)$ and $\overline{u}_{1,f}(N)$ are decoded. In channel use $N+1$, node~3/destination~1 decodes the desired F-signals $u_{1,f}(N-1)$ and $\overline{u}_{1,f}(N-1)$, in addition to its desired D-signal $u_{1,d}(N)$ (obtained from $u_{1,d}(N)\oplus u_{2,d}(N)$) and the C-signal sum $u_{1,c}(N)\oplus u_{2,c}(N)$. Next, in the $N$-th channel use, it decodes $u_{2,c}(N)$, $u_{1,f}(N-2)$, $\overline{u}_{1,f}(N-2)$, $u_{1,d}(N-1)$, and $u_{1,c}(N-1)\oplus u_{2,c}(N-1)$. Then it adds $u_{1,c}(N)\oplus u_{2,c}(N)$ to $u_{2,c}(N)$ to obtain the desired C-signal $u_{1,c}(N)$. Decoding proceeds backwards till channel use $i=1$. Similar processing is performed by node~4/destination~2. The number of bits recovered by node~3/destination~1 is $R_1=R_c+R_{1d}+R_{1f}+\overline{R}_f$, and similarly node~4/destination~2 obtains $R_2=R_c+R_{2d}+R_{2f}+\overline{R}_f$. 

\subsubsection{Achievable region}
Collecting the bounds \eqref{RCWFNsNc0}, \eqref{RCWFNsNc1}, \eqref{RCWFNsNc2}, \eqref{RCWFNsNc3a}, \eqref{RCWFNsNc3b}, and \eqref{RCWFNsNc4}, we see that a pair $(R_1,R_2)$ with $R_1=R_c+R_{1d}+R_{1f}+\overline{R}_f$ and $R_2=R_c+R_{2d}+R_{2f}+\overline{R}_f$, where the rates $R_c$, $R_{1d}$, $R_{2d}$, $R_{1f}$, $R_{2f}$, $\overline{R}_{f}$ are non-negative, is achievable if
\begin{align*}
R_{c}+R_{1d}+R_{2d}+R_{1f}+R_{2f}+\overline{R}_f&\leq n_s\\
R_{c}+R_{1d}+R_{2d}&\leq n_r\\
R_{1f}+\overline{R}_f&\leq n_f\\
R_{2f}+\overline{R}_f&\leq n_f\\
2R_{c}+2R_{1d}+2R_{2d}+R_{1f}+R_{2f}+2\overline{R}_f&\leq n_c.
\end{align*}
Solving this set of linear inequalities using the Fourier Motzkin elimination, we get the achievable region given by
\begin{align}
0\leq R_1&\leq\min\{n_s,n_r+n_f\}\\
0\leq R_2&\leq\min\{n_s,n_r+n_f\}\\
R_1+R_2&\leq n_c,
\end{align}
which proves Lemma \ref{Lemma:NrNs_Nc}. This also proves Thm.~\ref{Theorem:Capacity} for the case $\max\{n_r,n_s\}<n_c$.

\subsection{Case $n_r< n_c\leq n_s$}
\begin{figure}[ht]
\centering
\includegraphics[width=.7\columnwidth]{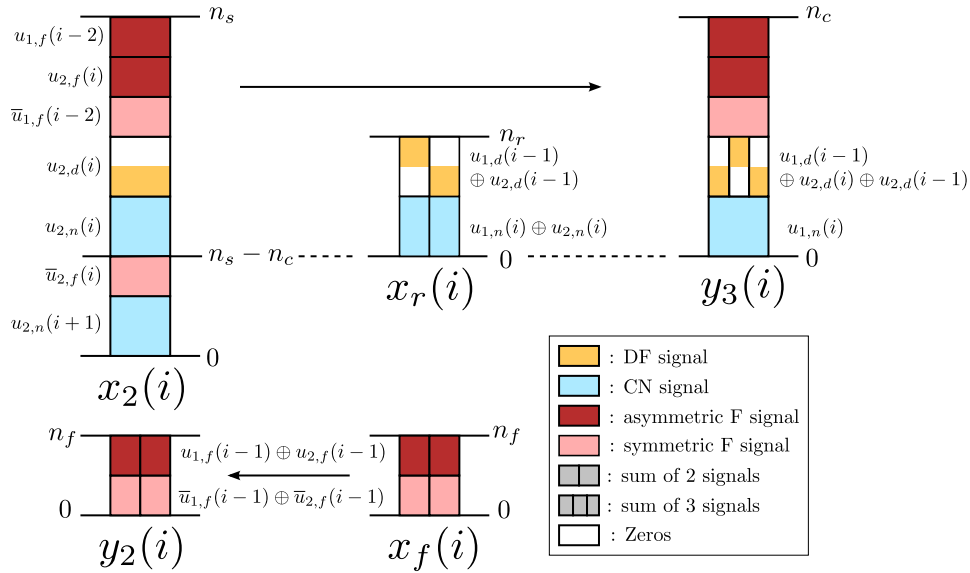}
\caption{The transmit signal and received signal of node~2, the transmit signals of the relay, and the received signal of node~3/destination~1 for the capacity achieving scheme of the linear deterministic BFN with feedback with $n_r< n_c\leq n_s$ and $n_f>0$. Node 2 makes use of the feedback signals $u_{1,f}(i-1)\oplus u_{2,f}(i-1)$ and $\overline{u}_{1,f}(i-1)\oplus \overline{u}_{2,f}(i-1)$ to extract $u_{1,f}(i-1)$ and $\overline{u}_{1,f}(i-1)$ which are sent to node~3/destination~1 in channel use $i+1$. Node~3/destination~1 decodes $u_{1,f}(i-2)$, $\overline{u}_{1,f}(i-2)$, $u_{1,d}(i-1)$, $u_{2,d}(i-1)$, and $u_{1,n}(i)$ in channel use $i$.}
\label{Fig:Scheme4}
\end{figure}

In this case, the relay observes more bits than the destinations since $n_s\geq n_c$. Thus, the sources can exploit the additional $n_s-n_c$ bits by using the CN strategy of Sect. \ref{Section:CodingStrategies:CN}. Additionally, we use the F strategy for feedback, and the DF strategy to achieve asymmetric rates. In the rest of this subsection, we prove the following lemma.
\begin{lemma}
\label{Lemma:Nr_Nc_Ns}
The region defined by
\begin{align*}
0\leq R_1&\leq \min\{n_r+n_f,n_c\}\\
0\leq R_2&\leq \min\{n_r+n_f,n_c\}\\
R_1+R_2&\leq n_s,
\end{align*}
is achievable in the linear deterministic BFN with feedback with  $n_r< n_c\leq n_s$.
\end{lemma}
This lemma proves Thm.~\ref{Theorem:Capacity} for the given case since the achievable region of this lemma matches the outer bound given in Thm.~\ref{Theorem:CapacityOuter}. Next, we describe the scheme which achieves the region in Lemma \ref{Lemma:Nr_Nc_Ns}. The transmit signals of node~2 and the relay, and the received signals at node~2 and node~3/destination~1 for the capacity achieving scheme are depicted graphically in Fig. \ref{Fig:Scheme4}.

\subsubsection{Encoding}
In this case, node~1 sends a D-signal vector $u_{1,d}(i)$ of length $R_{1d}+R_{2d}$ (zero padded as explained in Sect. \ref{Section:CodingStrategies:DF}), two N-signal vectors $u_{1,n}(i)$ and $u_{1,n}(i+1)$ of length $R_n$ each, two F-signal vectors $u_{1,f}(i)$ (asymmetric) and $\overline{u}_{1,f}(i)$ (symmetric) of length $R_{1f}$ and $\overline{R}_f$, respectively. Additionally, it sends the F-signals of node~2 (acquired through feedback) $u_{2,f}(i-2)$ and $\overline{u}_{2,f}(i-2)$ of length $R_{2f}$ and $\overline{R}_f$, respectively, as shown if Fig. \ref{Fig:Scheme4}.

Notice that out of these signals, two do not have to be observed at the destinations, namely $u_{1,n}(i+1)$ and $\overline{u}_{1,f}(i)$. These two signals have to be decoded at the relay to establish the F and the CN strategies. Thus, these signals can be sent below the noise floor of the destinations, i.e., in the lower $n_s-n_c$ levels observed at the relay. Assume that these signals do not fit in this interval of length $n_s-n_c$, i.e., $R_n+\overline{R}_f>n_s-n_c$. In this case, a part of these signals is sent below the noise floor, and a part above it. For this reason, we split these signals to two parts:
$$\overline{u}_{1,f}(i)=\left[\begin{array}{c}\overline{u}_{1,f}^{[1]}(i)\\\overline{u}_{1,f}^{[2]}(i)\end{array}\right],\quad u_{1,n}(i)=\left[\begin{array}{c}u_{1,n}^{[1]}(i)\\u_{1,n}^{[2]}(i)\end{array}\right],$$
where $\overline{u}_{1,f}^{[m]}$ has length $\overline{R}_f^{[m]}$ and $u_{1,n}^{[m]}$ has length $R_n^{[m]}$, $m=1,2$, such that $\overline{R}_f^{[1]}+\overline{R}_f^{[2]}=\overline{R}_f$ and $R_n^{[1]}+R_n^{[2]}=R_n$ (this split is not shown in Fig. \ref{Fig:Scheme4} for clarity). As a result, node~1 sends
\begin{align}
\label{Nr_Nc_Ns_X1}
x_1(i)=\left[\begin{array}{c} 
0_{n_c-R_{1f}-R_{2f}-2\overline{R}_f^{[1]}-\overline{R}_f^{[2]}-R_{1d}-R_{2d}-2R_n^{[1]}-R_n^{[2]}}\\
u_{1,f}(i)\\
u_{2,f}(i-2)\\
\overline{u}_{2,f}^{[1]}(i-2)\\
\overline{u}_{2,f}^{[2]}(i-2)\\
\overline{u}_{1,f}^{[1]}(i)\\
u_{1,n}^{[1]}(i+1)\\
u_{1,d}(i)\\
u_{1,n}^{[1]}(i)\\
u_{1,n}^{[2]}(i)\\
\overline{u}_{1,f}^{[2]}(i)\\
u_{1,n}^{[2]}(i+1)\\
0_{n_s-n_c-\overline{R}_f^{[2]}-R_n^{[2]}}\\
0_{q-n_s}
\end{array}\right].
\end{align}

The vectors $\overline{u}_{1,f}^{[1]}(i)$ and $u_{1,n}^{[1]}(i+1)$ are sent above $u_{1,d}(i)$, $u_{1,n}^{[1]}(i)$, and $u_{1,n}^{[2]}(i)$ since that latter signals have to align with the signals sent from the relay (see Sect. \ref{Section:CodingStrategies:CN} and \ref{Section:CodingStrategies:F}), where the relay can only access lower levels since $n_r<n_c$ in this case. The transmit signal of node~2, $x_2(i)$, is constructed similarly, by replacing $u_{1,f}(i)$ and $u_{2,f}(i-2)$ with $u_{1,f}(i-2)$ and $u_{2,f}(i)$, respectively, and replacing the user index of the other signals with 2. This construction requires
\begin{align}
\label{RCWFNcNs1}
R_{1f}+R_{2f}+2\overline{R}_f^{[1]}+\overline{R}_f^{[2]}+R_{1d}+R_{2d}+2R_n^{[1]}+R_n^{[2]}&\leq n_c\\
\label{RCWFNcNs2}
\overline{R}_f^{[2]}+R_n^{[2]}&\leq n_s-n_c.
\end{align}

\subsubsection{Relay processing}
\begin{figure}[ht]
\centering
\includegraphics[width=.75\columnwidth]{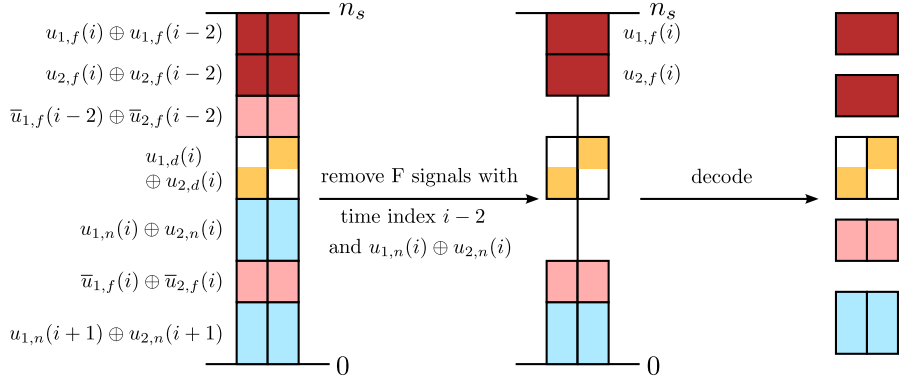}
\caption{The processing steps at the relay. The relay starts by removing its past-decoded signals. Then it decodes the signals $u_{1,f}(i)$, $u_{2,f}(i)$, $u_{1,d}(i)\oplus u_{2,d}(i)$, $\overline{u}_{1,f}(i)\oplus \overline{u}_{2,f}(i)$, and $u_{1,n}(i+1)\oplus u_{2,n}(i+1)$ in channel use $i$. In channel use $i+1$, the decoded F-signals are fed back to the sources, and the N-signal and D-signals are forwarded to the destinations.}
\label{Fig:Scheme4_Relay_Decoding}
\end{figure}

The relay receives the top-most $n_s$ bits of $x_1(i)\oplus x_2(i)$. We write $y_0(i)$ as 
\begin{align*}
y_0(i)=\left[\begin{array}{c}
0_{q-n_s}\\
0_{n_c-R_{1f}-R_{2f}-2\overline{R}_f^{[1]}-\overline{R}_f^{[2]}-R_{1d}-R_{2d}-2R_n^{[1]}-R_n^{[2]}}\\
u_{1,f}(i)\oplus u_{1,f}(i-2)\\
u_{2,f}(i)\oplus u_{2,f}(i-2)\\
\overline{u}_{1,f}^{[1]}(i-2)\oplus \overline{u}_{2,f}^{[1]}(i-2)\\
\overline{u}_{1,f}^{[2]}(i-2)\oplus \overline{u}_{2,f}^{[2]}(i-2)\\
\overline{u}_{1,f}^{[1]}(i)\oplus \overline{u}_{2,f}^{[1]}(i)\\
u_{1,n}^{[1]}(i+1)\oplus u_{2,f}^{[1]}(i+1)\\
u_{1,d}(i)\oplus u_{2,d}(i)\\
u_{1,n}^{[1]}(i)\oplus u_{2,f}^{[1]}(i)\\
u_{1,n}^{[2]}(i)\oplus u_{2,f}^{[2]}(i)\\
\overline{u}_{1,f}^{[2]}(i)\oplus \overline{u}_{2,f}^{[2]}(i)\\
u_{1,n}^{[2]}(i+1)\oplus u_{2,f}^{[2]}(i+1)\\
0_{n_s-R_{1f}-R_{2f}-2\overline{R}_f-R_{1d}-R_{2d}-2R_n}
\end{array}\right]
\end{align*}
The relay starts processing this signal by removing the past F-signals $u_{1,f}(i-2)$ and $u_{2,f}(i-2)$ (decoded in channel use $i-2$) from $y_0(i)$. Then it decodes the remaining signals as shown in Fig. \ref{Fig:Scheme4_Relay_Decoding}.

Then, the relay forwards 
\begin{align*}
x_{r}(i+1)=\left[\begin{array}{c} 
0_{n_r-R_{1d}-R_{2d}-R_{n}^{[1]}-R_{n}^{[2]}}\\
u_{1,d}(i)\oplus u_{2,d}(i)\\
u_{1,n}^{[1]}(i+1)\oplus u_{2,f}^{[1]}(i+1)\\
u_{1,n}^{[2]}(i+1)\oplus u_{2,f}^{[2]}(i+1)\\
0_{q-n_r}
\end{array}\right],
\end{align*}
in channel use $i+1$. The given signals fit in the interval of length $n_r$ if
\begin{align}
\label{RCWFNcNs3}
R_{1d}+R_{2d}+R_{n}^{[1]}+R_{n}^{[2]}\leq n_r.
\end{align}
The relay also sends a feedback signal $x_f(i+1)$ in channel use $i+1$ to node~1 and node~2 on the backward channel, where
\begin{align*}
x_{f}(i+1)=\left[\begin{array}{c} 
u_{1,f}(i)\oplus u_{2,f}(i)\\
\overline{u}_{1,f}^{[1]}(i)\oplus \overline{u}_{2,f}^{[1]}(i)\\
\overline{u}_{1,f}^{[2]}(i)\oplus \overline{u}_{2,f}^{[2]}(i)\\
0_{n_f-\max\{R_{1f},R_{2f}\}-\overline{R}_f^{[1]}-\overline{R}_f^{[2]}}\\
0_{q-n_f}
\end{array}\right].
\end{align*}
For efficient use of the feedback channel, the relay adds the signals $u_{1,f}(i)$ and $u_{2,f}(i)$ together, and feeds the sum back. If these signals do not have the same length, the shorter is zero padded at the relay till both signals have the same length. This construction requires
\begin{align}
\label{RCWFNcNs4}
R_{1f}+\overline{R}_f^{[1]}+\overline{R}_f^{[2]}&\leq n_f\\
\label{RCWFNcNs5}
R_{2f}+\overline{R}_f^{[1]}+\overline{R}_f^{[2]}&\leq n_f.
\end{align}

\subsubsection{Processing feedback at the sources}
Consider node~1 at channel use $i+1$. Node~1 receives the feedback signal
\begin{align*}
y_1(i+1)&=\left[\begin{array}{c} 
0_{q-n_f}\\
u_{1,f}(i)\oplus u_{2,f}(i)\\
\overline{u}_{1,f}^{[1]}(i)\oplus \overline{u}_{2,f}^{[1]}(i)\\
\overline{u}_{1,f}^{[2]}(i)\oplus \overline{u}_{2,f}^{[2]}(i)\\
0_{n_f-\max\{R_{1f},R_{2f}\}-\overline{R}_f^{[1]}-\overline{R}_f^{[2]}}\\
\end{array}\right].
\end{align*}
Node~1 then subtracts its own F-signals from $y_1(i+1)$, and obtains the F-signals of node~2, i.e., $u_{2,f}(i)$, $\overline{u}_{2,f}^{[1]}(i)$, and $\overline{u}_{2,f}^{[2]}(i)$. These signals are sent in channel use $i+2$ as seen in \eqref{Nr_Nc_Ns_X1}.

\subsubsection{Decoding at the destinations}
In the $i$-th channel use, node~3/destination~1 observes
\begin{align*}
y_{3}(i)=
\left[\begin{array}{c} 
0_{q-n_c}\\
0_{n_c-R_{1f}-R_{2f}-2\overline{R}_f^{[1]}-\overline{R}_f^{[2]}-R_{1d}-R_{2d}-2R_n^{[1]}-R_n^{[2]}}\\
u_{1,f}(i-2)\\
u_{2,f}(i)\\
\overline{u}_{1,f}^{[1]}(i-2)\\
\overline{u}_{1,f}^{[2]}(i-2)\\
\overline{u}_{2,f}^{[1]}(i)\\
u_{2,n}^{[1]}(i+1)\\
u_{2,d}(i)\oplus u_{1,d}(i-1)\oplus u_{2,d}(i-1)\\
u_{1,n}^{[1]}(i)\\
u_{1,n}^{[2]}(i)\\
\end{array}\right].
\end{align*}
Decoding at node~3/destination~1 is done in a backward fashion. In the $i$-th channel use, it starts with removing the already known D-signal $u_{2,d}(i)$ (decoded in channel use $i+1$). Then it proceeds with decoding each of 
$$u_{1,f}(i-2),\quad \overline{u}_{1,f}^{[1]}(i-2),\quad \overline{u}_{1,f}^{[2]}(i-2),\quad u_{1,d}(i-1),\quad u_{2,d}(i-1),\quad u_{1,n}^{[1]}(i),\quad u_{1,n}^{[2]}(i).$$ It recovers its desired signals for a total rate of $R_1=R_{1f}+\overline{R}_f^{[1]}+\overline{R}_f^{[2]}+R_{1d}+R_n^{[1]}+
R_n^{[2]}$. Similarly, node~4/destination~2 recovers $R_2=R_{2f}+\overline{R}_f^{[1]}+\overline{R}_f^{[2]}+R_{2d}+R_n^{[1]}+
R_n^{[2]}$ bits per channel use.

\subsubsection{Achievable region}
Collecting the bounds \eqref{RCWFNcNs1}, \eqref{RCWFNcNs2}, \eqref{RCWFNcNs3}, \eqref{RCWFNcNs4}, and \eqref{RCWFNcNs5} we get
\begin{align*}
R_{1f}+R_{2f}+2\overline{R}_f^{[1]}+\overline{R}_f^{[2]}+R_{1d}+R_{2d}+2R_n^{[1]}+R_n^{[2]}&\leq n_c\\
\overline{R}_f^{[2]}+R_n^{[2]}&\leq n_s-n_c\\
R_{1d}+R_{2d}+R_{n}^{[1]}+R_{n}^{[2]}&\leq n_r\\
R_{1f}+\overline{R}_f^{[1]}+\overline{R}_f^{[2]}&\leq n_f\\
R_{2f}+\overline{R}_f^{[1]}+\overline{R}_f^{[2]}&\leq n_f,
\end{align*}
where the rates $R_{1f}$, $R_{2f}$, $\overline{R}_{f}^{[1]}$, $\overline{R}_{f}^{[2]}$,
$R_{1d}$, $R_{2d}$, $R_{n}^{[1]}$, and $R_{n}^{[2]}$ are non-negative. Solving this set in linear inequalities using the Fourier Motzkin elimination with $R_1=R_{1f}+\overline{R}_f^{[1]}+\overline{R}_f^{[2]}+R_{1d}+R_n^{[1]}+
R_n^{[2]}$ and $R_2=R_{2f}+\overline{R}_f^{[1]}+\overline{R}_f^{[2]}+R_{2d}+R_n^{[1]}+
R_n^{[2]}$ yields the following achievable rate region
\begin{align}
0\leq R_1&\leq \min\{n_r+n_f,n_c\}\\
0\leq R_2&\leq \min\{n_r+n_f,n_c\}\\
R_1+R_2&\leq n_s,
\end{align}
which proves Lemma \ref{Lemma:Nr_Nc_Ns}. By the end of this section, we finish the proof of Thm.~\ref{Theorem:Capacity}.

\section{Net Feedback Gain}
\label{Section:NetGain}
At this point, it is clear that relay-source feedback link can increases the capacity of the BFN with respect to the non-feedback case. However, is this feedback efficient? In other words, is there a {\em net-gain} when using feedback?
In this section, we discuss the net-gain attained by exploiting feedback and we answer the question above in the affirmative. 

First, let us define what we mean by net-gain. Let $C_0$ be the sum-capacity of a BFN without feedback ($n_f=0$), and let $C_{n_f}$ be the sum-capacity with feedback ($n_f\neq0$), which is achieved by feeding back $r_f$ bits per channel use through the feedback channel. Let $\eta$ be defined as the ratio
\begin{align*}
\eta=\frac{C_{n_f}-C_0}{r_f}.
\end{align*}
We say that we have a net-gain if the ratio of the sum-capacity increase to the number of feedback bits is larger than 1, i.e., $\eta>1$. Otherwise, if $\eta\leq 1$, then we have no net-gain because then $C_{n_f}-C_0\leq r_f$, i.e., the gain is less than the number of bits sent over the feedback channel.

Note that if $n_c\leq n_r$, then there is no feedback gain at all, since in this case, the capacity region in Thm.~\ref{Theorem:Capacity} is the same as $n_f=0$. 

\begin{figure}[ht]
\centering
\includegraphics[width=.3\columnwidth]{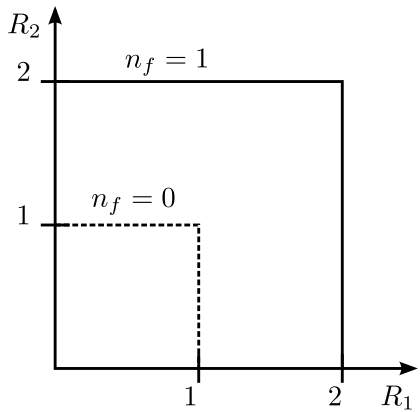}
\caption{The capacity region of the deterministic BFN with $(n_c,n_s,n_r)=(6,3,1)$ with ($n_f=1$) and without ($n_f=0$)  feedback.}
\label{Fig:NetGainRegion1}
\end{figure}

Now, consider for sake of example the case $n_c>n_f$ with a BFN with $(n_c,n_s,n_r)=(6,3,1)$. The capacity region of this BFN without feedback is shown in Fig.~\ref{Fig:NetGainRegion1}. 
The no-feedback sum-capacity of this network is $C_0=2$ bits per channel use corresponding to the rate pair $(R_1,R_2)=(1,1)$. This rate pair is achieved by using the CF strategy, where node~1 sends $x_1(i)=[u_{1,c}(i),\ 0_{5}^T]^T$ and node~2 sends $x_2(i)=[u_{2,c}(i),\ 0_{5}^T]^T$, and the relay sends $x_r(i)=[u_{1,c}(i-1)\oplus u_{2,c}(i-1),\ 0_{5}]^T$.
Now consider the case with $n_f=1$. In this case, the sum-capacity is $C_1=4$ bits per channel use corresponding to the corner point of the capacity region $(R_1,R_2)=(2,2)$ as shown in Fig.~\ref{Fig:NetGainRegion1}. To achieve this, the sources use the same CF strategy used for $n_f=0$, which achieves $R_1=R_2=1$ bit per channel use. Additionally each source sends a feedback bit $u_{j,f}(i)$ to the other source via the relay using the symmetric F strategy. This way, each source acquires the F-signal of the other source, which it forwards then to the respective destination. This F strategy requires feeding back only $r_f=1$ bit, namely $u_{1,f}(i)\oplus u_{2,f}(i)$. With this we have 
\[
\eta=\frac{C_1-C_0}{r_f}=\frac{4-2}{1}=2,
\]
i.e., a net-gain: {\em for each feedback bit, we gain 2 bits in the sum-capacity}.

\section{Summary}
\label{summary}
We have studied the butterfly network with relay-source feedback and examined the benefit of feedback for this network. We have derived capacity upper bounds, and proposed transmission schemes that exploit the feedback channel. The result was a characterization of the capacity region of the network. While feedback does not affect the capacity of the network in some cases, it does enlarge its capacity region in other cases. Moreover, the proposed feedback scheme which is based on bi-directional relaying is an efficient form of feedback, it provides a net-gain in the regimes where feedback helps. It turns out that the increase in the sum-capacity of the network is twice the number of feedback bits.

\section*{Acknowledgement}

The work of A. Chaaban and A. Segzin is supported by the German Research Foundation, Deutsche Forschungsgemeinschaft (DFG), Germany, under grant SE 1697/7.

The work of Dr. D.~Tuninetti was partially funded by NSF under award number 0643954;
the contents of this article are solely the responsibility of the author and
do not necessarily represent the official views of the NSF.
The work of Dr. D.~Tuninetti was possible thanks to the generous support of Telecom-ParisTech, Paris France,
while the author was on a sabbatical leave at the same institution.

\bibliography{myBib}

\begin{thebibliography}{10}
\providecommand{\url}[1]{#1}
\csname url@samestyle\endcsname
\providecommand{\newblock}{\relax}
\providecommand{\bibinfo}[2]{#2}
\providecommand{\BIBentrySTDinterwordspacing}{\spaceskip=0pt\relax}
\providecommand{\BIBentryALTinterwordstretchfactor}{4}
\providecommand{\BIBentryALTinterwordspacing}{\spaceskip=\fontdimen2\font plus
\BIBentryALTinterwordstretchfactor\fontdimen3\font minus
  \fontdimen4\font\relax}
\providecommand{\BIBforeignlanguage}[2]{{%
\expandafter\ifx\csname l@#1\endcsname\relax
\typeout{** WARNING: IEEEtran.bst: No hyphenation pattern has been}%
\typeout{** loaded for the language `#1'. Using the pattern for}%
\typeout{** the default language instead.}%
\else
\language=\csname l@#1\endcsname
\fi
#2}}
\providecommand{\BIBdecl}{\relax}
\BIBdecl

\bibitem{Ahlswede}
R.~Ahlswede, ``{Multi-way communication channels},'' in \emph{Proc. of 2nd
  International Symposium on Info. Theory}, Tsahkadsor, Armenian S.S.R., Sep.
  1971, pp. 23--52.

\bibitem{AvestimehrHo}
A.~S. Avestimehr and T.~Ho, ``{Approximate capacity of the symmetric
  half-duplex Gaussian butterfly network},'' in \emph{Proc. of the IEEE
  Information Theory Workshop (ITW)}, June 2009, pp. 311 -- 315.

\bibitem{AhlswedeCaiLiYeung}
R.~Ahlswede, N.~Cai, S.-Y.~R. Li, and R.~W. Yeung, ``{Network Information
  Flow},'' \emph{IEEE Trans. on Info. Theory}, vol.~46, no.~4, pp. 1204 --
  1216, July 2000.

\bibitem{AvestimehrDiggaviTse}
A.~S. Avestimehr, S.~Diggavi, and D.~Tse, ``{A deterministic approach to
  wireless relay networks},'' in \emph{Proc. of Allerton Conference}, 2007.

\bibitem{SahinErkip}
O.~Sahin and E.~Erkip, ``{Achievable rates for the Gaussian interference relay
  channel},'' in \emph{Proc. of 2007 GLOBECOM Communication Theory Symposium},
  Washington D.C., Nov. 2007.

\bibitem{MaricDaboraGoldsmith_IT}
I.~Mari\c{c}, R.~Dabora, and A.~J. Goldsmith, ``{Relaying in the Presence of
  Interference: Achievable Rates, Interference Forwarding, and Outer Bounds},''
  \emph{IEEE Trans. on Info. Theory}, vol.~58, no.~7, pp. 4342--4354, July
  2012.

\bibitem{ChaabanSezgin_IT_IRC}
A.~Chaaban and A.~Sezgin, ``{Achievable rates and upper bounds for the Gaussian
  interference relay channel},'' \emph{IEEE Trans. on Info. Theory}, vol.~58,
  no.~7, pp. 4432--4461, July 2012.

\bibitem{CoverThomas}
T.~Cover and J.~Thomas, \emph{{Elements of information theory (Second
  Edition)}}.\hskip 1em plus 0.5em minus 0.4em\relax John Wiley and Sons, Inc.,
  2006.

\bibitem{Tuninetti_ITW}
D.~Tuninetti, ``{An outer bound for the memoryless two-user interference
  channel with general cooperation},'' in \emph{proc. of the IEEE Information
  Theory Workshop (ITW)}, Lausanne, Switzerland, Sep. 2012.

\bibitem{CoverElgamal}
T.~M. Cover and A.~El-Gamal, ``{Capacity theorems for the relay channel},''
  \emph{IEEE Trans. on Info. Theory}, vol. IT-25, no.~5, pp. 572--584, Sep.
  1979.

\bibitem{NazerGastpar}
B.~Nazer and M.~Gastpar, ``{Compute-and-Forward: Harnessing interference
  through structured codes},'' \emph{IEEE Trans. on Info. Theory}, vol.~57,
  no.~10, pp. 6463 -- 6486, Oct. 2011.

\bibitem{YangTuninentti_Asilomar}
E.~Yang and D.~Tuninetti, ``{Interference channels with source cooperation in
  the strong cooperation regime: symmetric capacity to within 2 bits/s/Hz with
  dirty paper coding},'' in \emph{Proc. of 42nd Asilomar Conference on Signals,
  Systems and Computers}, Pacific Grove, CA, USA, Nov. 2011.

\bibitem{RankovWittneben}
B.~Rankov and A.~Wittneben, ``{Spectral efficient signaling for half-duplex
  relay channels},'' in \emph{Proc. of the Asilomar Conference on Signals,
  Systems, and Computers}, Pacific Grove, CA, Nov. 2005.

\bibitem{KimDevroyeMitranTarokh}
S.~Kim, N.~Devroye, P.~Mitran, and V.~Tarokh, ``{Comparisons of bi-directional
  relaying protocols},'' in \emph{Proc. of the IEEE Sarnoff Symposium},
  Princeton, NJ, Apr. 2008.

\bibitem{AvestimehrSezginTse}
A.~S. Avestimehr, A.Sezgin, and D.~Tse, ``{Capacity of the two-way relay
  channel within a constant gap},'' \emph{European Trans. in
  Telecommunications}, 2009.

\bibitem{Carleial}
A.~B. Carleial, ``{Interference channels},'' \emph{IEEE Trans. on Info.
  Theory}, vol.~24, no.~1, pp. 60--70, Jan. 1978.

\bibitem{HanKobayashi}
T.~S. Han and K.~Kobayashi, ``{A new achievable rate region for the
  interference channel},'' \emph{IEEE Trans. on Info. Theory}, vol. IT-27,
  no.~1, pp. 49--60, Jan. 1981.

\bibitem{NarayananPravinSprintson}
K.~Narayanan, M.~P. Wilson, and A.~Sprintson, ``{Joint physical layer coding
  and network coding for bi-directional relaying},'' in \emph{Proc. of the
  Forty-Fifth Allerton Conference}, Illinois, USA, Sep. 2007.

\bibitem{ElgamalKim}
A.~E. Gamal and Y.-H. Kim, \emph{{Network Information Theory}}.\hskip 1em plus
  0.5em minus 0.4em\relax Cambridge University Press, 2011.

\end{thebibliography}

\appendix
\section{Proof of Thm.~\ref{Theorem:Capacity}}

We set $X_0=(X_r,X_f)$ and use the output definition in~\eqref{eq:finallythemodelwestudy} in the outer bounds in Section~\ref{DMIRCWF:upper}.

From the cut-set bound in \eqref{CutSetBound1} we have
\begin{align*}
R_1
&\leq I(X_1;Y_0,Y_2,Y_3|X_0,X_2)\\
&= I(X_1;Y_0,Y_2,Y_3|X_r,X_f,X_2)\\
&= H(Y_0,Y_2,Y_3|X_r,X_f,X_2)-H(Y_0,Y_2,Y_3|X_r,X_f,X_2,X_1)\\
&= H(\mathbf{S}^{q-n_s}X_1|X_r,X_f,X_2)\\
&\leq H(\mathbf{S}^{q-n_s}X_1)\\
&\leq n_s,
\end{align*}
Similarly, the cut-set bounds in~\eqref{CutSetBound3} and~\eqref{CutSetBound4} reduce to
\begin{align*}
R_1&\leq n_r+n_f,\\
R_1&\leq \max\{n_c,n_r\},
\end{align*}
respectively.
These bounds combined give~\eqref{UpperBound1}. Similarly, the bound in~\eqref{UpperBound2} for $R_2$ follows by the symmetry in the network.

The sum-rate cut-set bound in~\eqref{CutSetBound6} becomes
\begin{align*}
R_1+R_2
&\leq I(X_0,X_1,X_2;Y_3,Y_4)\\
&= I(X_r,X_f,X_1,X_2;Y_3,Y_4)\\
&= H(Y_3,Y_4)\\
&= H(Y_3)+H(Y_4|Y_3),
\end{align*}
which leads to
\begin{align}
R_1+R_2 &\leq \max\{n_r,n_c\}+n_c. \label{CutSetBound6-d}
\end{align}
These are the neccessary cut-set upper bounds for our problem. The remaining cut-set bounds are redundant given the cooperation bounds that we derive next, and are thus omitted.

Next, we evaluate the cooperation bounds in~\eqref{eq:tuninetti itw 2012 lausanne allall}. In the symmetric case, bounds \eqref{eq:tuninetti itw 2012 lausanne 2-1}, \eqref{eq:tuninetti itw 2012 lausanne 2-2}, \eqref{eq:tuninetti itw 2012 lausanne 4-1},  and~\eqref{eq:tuninetti itw 2012 lausanne 4-2} are equivalent to 
bounds~\eqref{eq:tuninetti itw 2012 lausanne 1-2}, \eqref{eq:tuninetti itw 2012 lausanne 1-1}, \eqref{eq:tuninetti itw 2012 lausanne 3-2}, and~\eqref{eq:tuninetti itw 2012 lausanne 3-1}, respectively. Notice also that due to symmetry, the bounds \eqref{eq:tuninetti itw 2012 lausanne 1-2} and \eqref{eq:tuninetti itw 2012 lausanne 3-1} are similar. Thus, we need only to specialize the bounds \eqref{eq:tuninetti itw 2012 lausanne 1-1}, \eqref{eq:tuninetti itw 2012 lausanne 1-2}, and \eqref{eq:tuninetti itw 2012 lausanne 3-2} to the linear deterministic BFN with feedback. It turns out that the bound~\eqref{eq:tuninetti itw 2012 lausanne 1-1} for the linear deterministic BFN with feedback is redundant given~\eqref{CutSetBound6-d}. Thus, we omit its derivation. 

Next, we consider the bound in~\eqref{eq:tuninetti itw 2012 lausanne 1-2}, which yields
\begin{align*}
R_1+R_2  
  &\leq  I(X_2;Y_4,Y_1,Y_0| Y_3,X_1,X_0)
       + I(X_1,X_0,X_2;Y_3) 
\\&=  H(Y_4,Y_1,Y_0| Y_3,X_1,X_r,X_f)-H(Y_4,Y_1,Y_0| Y_3,X_1,X_r,X_f,X_2)\\
&\quad       + H(Y_3) -H(Y_3|X_1,X_0,X_2)
\\&=  H(\mathbf{S}^{q-n_{s}}X_2| \mathbf{S}^{q-n_{c}}X_2,X_1,X_r,X_f)
       + H(\mathbf{S}^{q-n_{c}}X_2+\mathbf{S}^{q-n_{r}}X_r)
\\&\leq (n_{s}-n_{c})^+ + \max\{n_{c},n_{r}\}.
\end{align*}
Notice that this bound can be tighter than the sum-rate cut-set bound in~\eqref{CutSetBound6-d} and is equal to~\eqref{UpperBound4}.

Finally, the bound in~\eqref{eq:tuninetti itw 2012 lausanne 3-2} becomes
\begin{align*}
R_1+R_2
  &\leq  I(X_2;Y_4,Y_1| Y_3,Y_0,X_1,X_0)
       + I(X_1,X_2;Y_3,Y_0|X_0)
\\&=     H(Y_4,Y_1| Y_3,Y_0,X_1,X_r,X_f)-H(Y_4,Y_1| Y_3,Y_0,X_1,X_r,X_f,X_2)
\\&   \quad    + H(Y_3,Y_0|X_r,X_f)-H(Y_3,Y_0|X_r,X_f,X_1,X_2)
\\&=     H(\mathbf{S}^{q-n_{c}}X_2,\mathbf{S}^{q-n_{s}}X_1+\mathbf{S}^{q-n_{s}}X_2|X_r,X_f)  
\\&\leq n_{s}+n_{c}.
\end{align*}
This bound yields~\eqref{UpperBound5}.

\end{document}